\documentclass[
 reprint,
 amsmath,
 amssymb,
aip]{revtex4-1}

\usepackage{graphicx}
\usepackage{dcolumn}
\usepackage{bm}

\usepackage{xcolor}
\usepackage{colortbl}
\usepackage{color}
\usepackage{tikz}
\usepackage{hyperref}
\hypersetup{pdfpagemode=FullScreen}
\usepackage{ifthen}
\usepackage{pgfplots}
\pgfplotsset{width=10cm,compat=1.9}
\definecolor{mma1}{rgb}{0.3725,0.5098,0.7020}
\definecolor{mma2}{rgb}{0.8745,0.6078,0.2039}
\definecolor{mma3}{rgb}{0.507813,0.714844,0.2039}
\definecolor{mma4}{rgb}{0.9137,0.3882,0.2398}
\usepackage{longtable}

\begin{document}

\title{
Freezing molecules with light: How long can one maintain a non-equilibrium
molecular geometry by strong light-matter coupling?}

\author{Eric R. Bittner}
\email{ebittner@central.uh.edu}
\affiliation{Department of Chemistry,
University of Houston, Houston, TX 77204}

\author{Ravyn A. Malatesta}
\affiliation{School of Chemistry and Biochemistry, Georgia Institute of Technology, 901 Atlantic Drive, Atlanta, GA~30332}

\author{Gabrielle D. Olinger}
\affiliation{Department of Physics,
University of Houston, Houston, TX 77204}

\author{Carlos~Silva-Acu\~na}
\affiliation{School of Chemistry and Biochemistry, Georgia Institute of Technology, 901 Atlantic Drive, Atlanta, GA~30332}
\affiliation{School of Physics, Georgia Institute of Technology, 837 State Street, Atlanta, GA~30332}

\date{\today}

\begin{abstract}
In molecular photochemistry, the non-equilibrium character and subsequent ultrafast relaxation dynamics of photoexcitations near the Franck-Condon region limit the control of their chemical reactivity.We address how to harness strong light-matter coupling in optical microcavities 
to isolate and preferentially select specific reaction pathways out of the myriad of possibilities present in large-scale complex systems.  
Using Fermi's Golden Rule and realistic molecular parameters, we estimate the extent to which molecular configurations can be ``locked" into non-equilibrium excited state configurations for timescales well beyond their natural relaxation times.  For upper polaritons--which are largely excitonic in character, molecular systems can be locked into their ground state geometries for tens to thousands of picoseconds and varies with the strength of the exciton/phonon coupling (Huang-Rhys parameter). On the other hand, relaxed LP lifetimes are nearly uniformly distributed between 2.1 -- 2.4\,ps and are nearly independent of the Huang-Rhys parameter. 
\end{abstract}


\maketitle

\section{Introduction}

The past few years has seen a remarkable 
surge in the use and application of photonic optical 
cavities to induce, manipulate, and modify chemical
process through strong coupling between quantized cavity 
photons and molecular states to form polaritons.
\cite{Eizner:2016aa,Feist:2018aa,Flick:2017aa,Galego:2015aa,Galego:2016aa,Galego:2017aa}
Cavity polaritons arise from non-perturbative coupling between the strongly quantized modes of the electromagnetic fields in the optical cavity and the optical transitions of the molecular species within the cavity.  These transitions can be vibrational or vibronic depending upon the tuning of the cavity.  

In the absence of a cavity, the electronic transitions can described within a Franck-Condon description  whereby optical transitions from the ground state to the excited state occur within a fixed nuclear frame;
this is followed by relaxation/reorganization on the excited state potential to some new local minimum energy geometry and then fluorescence or non-radiative decay carries the system back to the ground state.  Generally, the reorganization occurs on a timescale on the order of ps, fluorescence occurs on the order of ns, and non-radiative decay occurs on timescales the order of ns to ms.  As a result, fluorescence occurs according to Kasha's rule from the lowest lying vibronic state of the excited state potential to the ground state potential that gives rise to the vibronic fine structure readily observed these systems.  There may also be other important intersystem processes competing with this, such as charge-transfer and other photophysical processes that are initiated by the relaxation from the Franck-Condon point.

\begin{figure}
    \centering
    \includegraphics[width=\columnwidth]{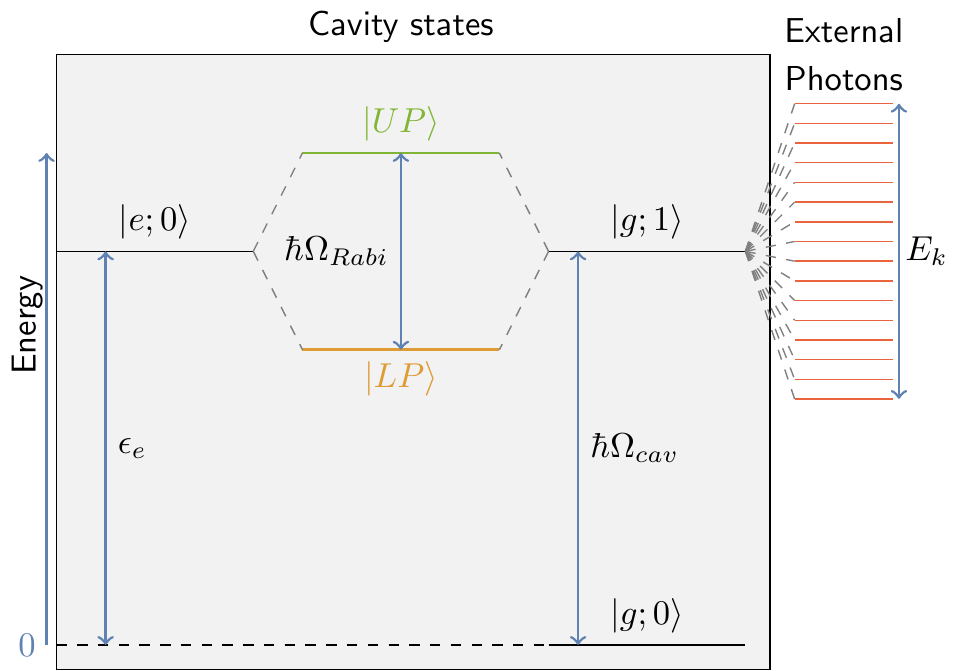}
    \caption{Energy level diagram depicting the electronic and photon excitations in the microcavity. }
    \label{fig:dressed}
\end{figure}

\begin{figure}
    \centering
\includegraphics[width=0.95\columnwidth]{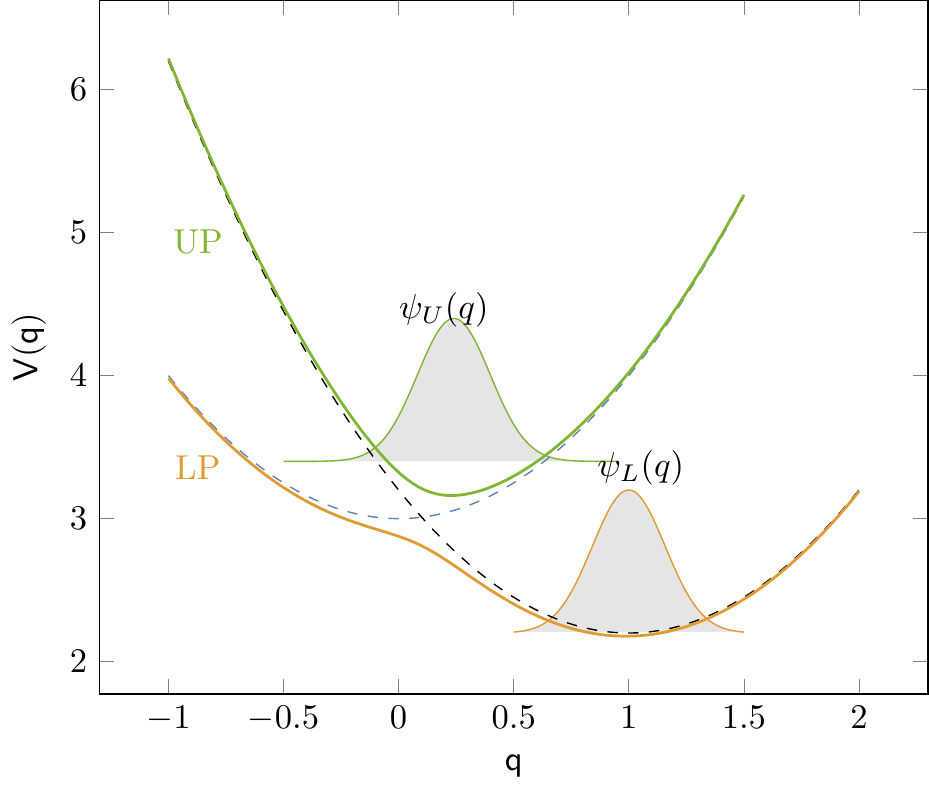}
    \caption{Schematic potential surfaces for 
    a molecular system interacting with a 
    single-mode cavity.}
    \label{fig:poten}
\end{figure}

In a microcavity, this picture needs to be modified in number 
of  important ways.\cite{Agranovich:2003aa,Herrera:2016aa,Herrera:2017aa,Herrera:2017ab,Herrera:2018aa}  First, the correct quantum mechanics basis  includes the quantized radiation field states into the atomic or molecular basis.  We denote these as $|\phi;n_k\rangle$ where $\phi$ denotes the material (atomic, molecular, etc.) state while $n_k$ denotes the number of quanta (photons) in the $k$-th mode of the radiation field.  Thus, a molecule in the ground state with a cavity devoid of photons would be denoted as $|g;0_k\rangle$ with energy $E_g+ \hbar\omega_k$.  This is depicted in Fig.\ref{fig:poten} in which to a first approximation the electronic potential surfaces are vertically shifted in energy by $\hbar\omega_kn_k$. For clarity, we show the case for a single mode cavity with frequency $\omega_c$. If the cavity is resonant or nearly resonant with the optical transition frequency, $\omega_x$, the adiabatic potential for the $|g;n_k\rangle$ 
(\textit{i.e.} molecular ground state + $n_k$ cavity photons) intersects the potential for the $|e;(n_k-1)\rangle$ (\textit{i.e.} molecular excited state + $n_k-1$ cavity photons) near the Franck-Condon point. Coupling between the radiation field and the atomic or molecule state, 
introduces a new set of eigenstates of the matter/radiation field which are linear combinations of the dressed-atom states termed lower (LP) and upper (UP) polaritons. 
\begin{align}
\left(
\begin{array}{c}
    |L_k(n_k)\rangle \\
    |U_k(n_k)\rangle 
    \end{array}
\right)
=
\left[
\begin{array}{cc}
c_k & s_k\\
-s_k & c_k
\end{array}
\right]
\left(
\begin{array}{c}
|g;n_k\rangle \\|e;(n_k-1)\rangle
\end{array}
\right)
\end{align}
with coefficients $c^2_k  + s^2_k = 1$.
At the crossing point, the two 
states are split by the Rabi frequency
$\hbar\Omega_{Rabi} = 2 |\mu_{eg}|\sqrt{n_k}E_{k}$
where  $E_{k} =\sqrt{\hbar\omega_{k}/2\epsilon V_{cav}}$
is the electric field of the cavity (assuming
optimal alignment between the 
molecular transition moments and the electric 
field in the cavity and cavity permittivity 
$\epsilon$ and volume $V_{cav}$). 

A second but equally important consideration is that the 
eigenstates of the cavity (the polariton states) are embedded 
in a continuum of photon states 
exterior to the cavity which ``dress''
the cavity states.\cite{Cohen-Tannoudji:aa}
Within this picture, sketched in Fig.~\ref{fig:dressed}, each polariton state, $\phi$,
is
embedded into the continuum by writing it as
\begin{align}
    |\psi_\mu\rangle = \sum_\phi c_\phi|\phi\rangle
    + \sum_{n}c_n|n\rangle.
    \label{eq:dressed}
\end{align}
These are formally eigenstates 
states of a Schr\"odinger equation
$(H_{cav} + H_{ext} + \hat W)|\psi_\mu\rangle = E_\mu |\psi_\mu\rangle$ 
where $H_{cav}$ is the cavity Hamiltonian 
 with eigenstates $|\phi\rangle$
with energies $E_\phi$.
$H_{ext}$ is the Hamiltonian 
for a quasi-continuum of  external states 
$|n\rangle$ with energy 
$E_n = n \delta$, respectively.  Here, integer $n\in (-\infty,\infty)$ indexes the quasi-continuum states 
with uniform spacing $\delta$.  Our final results 
will require $\delta \to 0$. 

We write  $\hat W$ as the coupling between 
the internal (cavity) and external states. 
An initially prepared  
interior (cavity) state will 
decay exponentially into the quasi-continuum 
of external states with time constant 
$\Gamma_{\phi}$ as per Fermi's golden rule
(FGR)
\begin{align}
    \Gamma_\phi = \frac{2\pi}{\hbar}
    |\langle\psi_\mu |\hat W | \phi\rangle |^2\frac{1}{\delta} 
\end{align}
where $\delta \to 0$ is the spacing between the
quasi-continuum states exterior to the cavity. 

Ordinarily, the UP and LP decay rates
are ``inherited" from the properties of the 
cavity and can be written as a weighted sum 
of the exciton decay rate and the cavity leakage
rate
\begin{align}
    \Gamma_{lp} = |s_k|^2\Gamma_{ex} + |c_k|^2\Gamma_{cav}
    \label{eq:4}
   \end{align}
This implies that the lower polariton lifetime
is set by the fastest decaying component 
which is generally the lifetime 
of a photon in the cavity.   However, 
experimental studies of organic microcavity 
systems suggest that their 
polariton states are much longer lived.\cite{Mony:2018aa,Wang:2014vr,Ballarini:2014tw}
This can be explained within the
exciton bath-model 
developed by Lidsey et al, 
the kinetics of populating the 
lower (emissive) polariton state 
is slow compared to its emission rate.
\cite{PhysRevB.65.195312}
Furthermore, the observed emission 
rates are not angle dependent, suggesting 
that there is sufficient thermalization of the 
cavity states.

Of central concern of 
this paper is the 
cavity induced (or suppressed) dynamics
of the molecular species on the UL and LP adiabatic
potentials and importantly, the non-adiabatic 
coupling between the two.  Vibrational and nuclear
motions about and away from this point introduce coupling  
between the UP and LP states. 
In other words, the light-matter
coupling introduces a non-adiabatic coupling between UP and LP adiabatic potentials. 
This non-adiabatic coupling affects the 
lifetime of molecular polaritons.
The relaxed exciton is decoupled from the 
cavity and population in this state serves as
a ``dark reservior''. 

In this paper we use Fermi's golden rule (FGR)
to compute the both UP and LP lifetimes
using a vibronic coupling model.  
First, we will work with a exciton/phonon model 
that incorporates on- and off-diagonal couplings between 
the electronic terms and the internal vibrational modes
of the molecular system.  We append to this 
the coupling between the electronic terms and 
the cavity photon modes within the Tavis-Cummings model. 
We use this as the starting point for computing the 
non-adiabatic coupling between the UP and LP polariton 
adiabatic potentials due to the vibronic motions of the
molecule on these potentials.  
We arrive at simple expressions based upon the molecular
vibronic couplings and frequencies and use this to 
estimate UP to LP conversion times for a wide range
of molecular chromophore systems.  

\section{Vibronic coupling between polariton branches}

To frame our discussion, we use the following
exciton/phonon model Hamiltonian
\begin{align}
 H_{ex}  &=    \sum_i\epsilon_i|i\rangle\langle i|
+\sum_{ijq}g_{ijq}(|i\rangle\langle  j|+|j\rangle\langle i|)(\hat b_q^\dagger +\hat b_q)
\nonumber \\
&+\sum_q \hbar\omega_q \hat b^\dagger_q \hat b_q
\label{eq:1}
\end{align}
where $\epsilon_i$ are the electronic energies 
referenced to common 
molecular geometry (e.g. the ground state, $i=0$, and excited states $i>0$), 
$g_{ijq}$ are the diagonal and 
off-diagonal derivatives of the potential at this geometry, and $[\hat b_q,\hat b_{q'}^\dagger]=\delta_{qq'}$ are boson operators for vibrational phonon
modes with frequency $\omega_q$.  
Each of these parameters can be 
determined by quantum chemical means. \cite{thouin2019polaron,yang2014intramolecular,yang2015computing,yang2017identifying,yang2018inelastic,karabunarliev:3988,karabunarliev:4291,karabunarliev:5863,karabunarliev:10219,karabunarliev:057402,karabunarliev:079901}
Importantly, the $\epsilon_i$ energies correspond to
the electronic eigenenergies {\em defined} at 
some local energy minimum. Correspondingly, 
the phonon normal modes, frequencies $\hbar\omega_q$, 
and coupling forces $g_{ijq}$ are similarly defined
relative to this configuration.  
We append to this the 
radiation modes of a microcavity and couplings within the Tavis/Cummings model\cite{Tavis:1969aa}
\begin{align}
    \hat V_{cav} = \sum_{k}\hbar\Omega_{k,cav} 
    \hat a^\dagger_k \hat a_k 
    + \sum_{k,i>0}
    \mu_{i0}E_o(|0\rangle\langle i |  \hat a^\dagger_k + |i\rangle\langle 0| \hat a_k)
    \label{eq:2}
\end{align}
where we take $\mu_{i0}E_o$ to be the projection of the
transition moment between the ground and excited states onto the 
electric field in the cavity. 
We assume the cavity has a single mode that is nearly resonant with a single electronic transition and work within a dressed-atom
basis and write $|0\rangle = |g,1_k\rangle$ as the
electronic ground-state with $1_k$ cavity
photons and $|1\rangle = |e,0_k\rangle$ as the first exciton state with $0_k$ cavity photons.

Fig.\ref{fig:poten} is a sketch of the
adiabatic potentials for molecular
motions in the  UP and LP eigenbasis. 
The dashed (diabatic) potentials
correspond to the diabatic potentials for the 
ground and excited molecular states; however, the
ground state potential is shifted up by the
energy of a single cavity photon.  In the scenario
shown here, the cavity is resonant with the 
absorption frequency of the molecule. 
Vibrational wavepacket dynamics then follow 
the lower (orange) adiabatic potential
away from the Franck-Condon point.  
As the molecule distorts along coordinate $q$, 
the resulting polariton state evolves
from being a 1:1 linear combination of 
a cavity photon and molecular excitation to 
being purely excitonic by the time it has 
relaxed.  

\begin{widetext}
Transforming to the UP/LP basis and assuming the 
cavity contains a single excitation, the cavity
Hamiltonian takes the form
\begin{align}
    \tilde H_{cav} &= H_{ex} + V_{cav} \nonumber \\
    &=
    \left(
    \begin{array}{cc}
        L_k^\dagger, &  U_k^\dagger
    \end{array}
    \right)
    .
    \left[
    \begin{array}{cc}
        \hbar\Omega_{L,k} + (s^2_kg_1-2 c_ks_k g_{01})\cdot X & (g_{01}(c_k^2-s_k^2)-c_ks_k g_1)\cdot X \\
        (g_{01}(c_k^2-s_k^2)-c_ks_k g_1)\cdot X  & \hbar\Omega_{U,k}+(c^2_kg_1+2 c_ks_k g_{01})\cdot X 
    \end{array}
    \right]
    \cdot
    \left(
    \begin{array}{c}
       \hat L_k ,\\ \hat U_k
    \end{array}
    \right)
    +\hbar\Omega \hat B^\dagger\hat B \nonumber \\
        &=
    \left(
    \begin{array}{cc}
        \hat L_k^\dagger ,&  \hat U_k^\dagger
    \end{array}
    \right)
    \left[
    \left[
    \begin{array}{cc}
        \hbar\Omega_{L,k} & 0 \\
        0  & \hbar\Omega_{U,k} 
    \end{array}
    \right]
+    \left[
    \begin{array}{cc}
        G_L & G_{LU} \\
        G_{UL}  &G_U 
    \end{array}
    \right]\cdot(B+B^\dagger)
    \right]
    \left(
    \begin{array}{c}
        \hat L_k \\ \hat U_k
    \end{array}
    \right)
    +\hbar\Omega \hat B^\dagger \hat B
    \label{eq:h-full},
\end{align}
where the coefficients $G_L$, $G_U$, and
$G_{UL}$ contain contributions from
both the cavity and the original (cavity free) electron/phonon coupling. The $G_L$ and $G_U$ 
will introduce a reorganization energy to 
both the UP and LP potential surfaces while the
$G_{LU}$ and $G_{UL}$ terms couple the UP and LP
polariton branches via both diagonal and off-diagonal exciton/phonon coupling terms. 
This generalized model can be parameterized 
from  {\em ab initio} calculations.  
\end{widetext}
\subsection{Reduced Model}
To analyse the interplay between 
the internal degrees 
of freedom and the cavity,  
we consider a far simpler model
of a two-electronic state molecule 
with a single vibrational degree of freedom
with frequency $\omega$
\begin{align}
    H_{ex} &=  \epsilon_g |g\rangle\langle g|
    +|e\rangle\langle e|\left(\epsilon_e + \hbar\omega S (b^\dagger + b)\right) \nonumber 
    \\
    &+g_{eg}(|g\rangle\langle e| +|e\rangle \langle g|)(b^\dagger + b)
    +\hbar\omega b^\dagger b.
    \label{eq:h-red}
\end{align}
Here, $S$ is a dimensionless (Huang-Rhys) parameter that determines the 
relative shift between the ground state potential and the 
excited state diabatic potentials.  
Under this simplified model, 
\begin{align}
    G_L &= \sin(\eta)^2 \hbar\omega S - 2 g_{eg}\sin(2 \eta) \\
    G_U &= \cos(\eta)^2 \hbar\omega S + 2 g_{eg}\sin(2 \eta) \\
    G_{LU}&=  g_{eg}\cos(2\eta) -\hbar \omega S \sin(2 \eta)/2
\end{align}
where
\begin{align}
    \frac{1}{2}\tan(2\eta) = -\frac{\lambda_{eg}}{(\epsilon_e-\epsilon_g)-\hbar\Omega_{cav}}
\end{align}
defines the mixing between the electronic transition
between $|g\rangle$ and $|e\rangle$
and the photon mode of the cavity and $\lambda_{eg} = \mu_{eg}E_o\sqrt{n_k}$
includes the molecular transition moment, field, and mode occupancy $n_k$.


\subsection{Embedding into the continuum}
Thus far, we have neglected the dressing
of the cavity states (i.e. the UP and LP polaritons)
by their coupling $\hat W$ to the photon states 
which are external to the cavity. 
For this, we apply the approach 
developed in Ref ~\citenum{Cohen-Tannoudji:aa}
whereby we write $|\psi_\mu\rangle$ as a 
solution of the full cavity + continuum 
Schr\"odinger equation
\begin{align}
(\hat H_{cav} + H_{ext} + \hat W) |\psi_{\mu}\rangle = E_\mu |\psi_\mu\rangle
\end{align}
Accordingly, we define the following matrix
elements:
\begin{align}
    v_n = \langle n | \hat W | LP \rangle  = v\\
    v'_n = \langle n | \hat W | UP \rangle  = v'\\
    0 = \langle n | \hat W| n' \rangle  = \langle n | H_{ext}| LP \rangle = \langle n | H_{ext}| UP \rangle 
\end{align}
Direct (phonon mediated) 
coupling between the UP and LP is mediated
by the off-diagonal terms in 
Eq.\ref{eq:h-full} and Eq.\ref{eq:h-red}.
For compactness in notation 
and to parallel the development in Ref.\cite{Cohen-Tannoudji:aa}
we shall write 
\begin{align}
    \hat w = G_{LU}(b^\dagger + b)
\end{align}
and recognize that we will need to average over 
the initial vibrational states to obtain our final expressions.

\subsection{Why Eq.~\ref{eq:4} is incomplete}
The assumption that the 
polariton decay is dictated by the
cavity lifetime can be analyzed 
using the FGR approach. 
However, rather than using the 
vibronic coupling (as suggested above) 
as the perturbation, 
we take the light-matter interaction as 
the perturbation and consider the 
decay of the molecular excitation $|e;0\rangle$ into the quasi-continuum.
The FGR rate for the exciton
decay is then given by
\begin{align}
\Gamma_{ex} = \frac{2\pi}{\hbar}|\langle \psi_\mu|
\hat W | e;0\rangle|^2 \frac{1}{\delta}.
\end{align}
If we assume that the decay
of $|e;0\rangle$ is via 
$|g;1\rangle$, which in turn can decay to
$|g;0\rangle$ by 
photon leakage from 
the cavity, then the coupling between 
the exciton and the continuum is given by
\begin{align}
\langle \psi_\mu|
\hat W | e;0\rangle = \langle \psi_\mu|
g;1\rangle \langle g;1| \hat W | e;0\rangle
\end{align}
Thus, the rate depends upon the cavity detuning.
\begin{align}
    \Gamma_{ex} = |\lambda_{eg}|^2 \frac{\Gamma_{cav}}{(\hbar \Gamma_{cav}/2)^2+(\epsilon_e - \hbar\omega_{cav})^2}.
\label{eq:19}
\end{align}
For strongly detuned cavities in which $|\epsilon_e -\hbar\omega_cav| \gg \hbar\Gamma_{cav}$, the
exciton can not decay by radiating into the continuum.
For a resonant cavity, on the other hand, 
\begin{align}
    \Gamma_{ex,res} = \frac{\Omega_{Rabi}^2}{\Gamma_{cav}}.
    \label{eq:20}
\end{align}
However, this expression is not valid
if the Rabi frequency is large
compared to $\Gamma_{cav}$, which is the energy width over which 
the coupling between the exciton and the continuum is significant. 
That is to say, that the cavity lifetime
must be significantly shorter than the 
Rabi oscillation period in order for Eq.~\ref{eq:4}  to be valid.

\subsection{UP Decay}
The probability amplitude that a continuum state $|\psi_\mu\rangle$
will be excited from the UP is proportional to the square of the 
following
matrix element:
\begin{align}
    \langle \psi_\mu| \hat W | UP\rangle
    &=\frac{\hat wv + v'\sum_nv^2/(E_\mu-E_n)}{(v^2+(\hbar\Gamma_{cav}/2)^2+E_\mu^2)^{1/2}}
    \nonumber\\
    &=
    \frac{\hat wv + v'E_\mu}{(v^2+(\hbar\Gamma_{cav}/2)^2+E_\mu^2)^{1/2}}.
    \label{eq:fano}
\end{align}
where we recall that $\hat w$ is the phonon-mediated
UP/LP coupling, and $v$ and $v'$ are the 
couplings between the LP and UP polaritons to the 
external photon states, respectively.  
The FGR expression is obtained by taking the 
trace over the vibrational degrees of freedom.
and gives an expression of the form
\begin{align}
    \Gamma_U = \Gamma_{U,direct} + \Gamma_{U,seq}
\end{align}
The first term gives the direct rate resulting from 
the decay of the UP directly
into the continuum via the $v'$ coupling
\begin{align}
\Gamma_{U,direct} =  \frac{2\pi}{\hbar}\frac{(v')^2}{\delta} \frac{E_\mu^2}{v^2 + (\hbar \Gamma_{cav}/2)^2 + E_\mu^2)}
\end{align}
which reduces to Eqs.~\ref{eq:19} and \ref{eq:20}
upon taking $\delta\to 0$ and 
setting $v' = v$.
The second  term arises from the fact that the UP 
can first decay to the LP, which in turn is also
embedded in the continuum via the sequential pathway
\begin{align} 
|UP\rangle \xrightarrow{\hat w} |LP \rangle \xrightarrow{ v} |k\rangle
\end{align}
If assume 
that the sequential path dominates
so that the relaxation is sequential, 
the matrix element between $UP$ and 
the continuum states factors into 
\begin{align}
    \langle \psi_\mu | W | {UP} \rangle
    &= \langle \psi_\mu|LP\rangle \langle LP | \hat W | UP \rangle  = \langle \psi_\mu|LP\rangle \hat w.
\end{align}
This  gives a FGR expression
    \begin{align}
\Gamma_{U,seq}&= 
\langle \hat w^2\rangle_{th}\frac{\Gamma_{cav}}{(\hbar\Gamma_{cav}/2)^2 +(\hbar\Omega_{Rabi})^2} 
\end{align}
The matrix element $\langle w^2\rangle_{th}$ contains contribution
from both the cavity and the internal vibronic coupling. If 
we average over a distribution of initial vibronic states
\begin{align}
    \langle \hat w^2\rangle_{th}
     \approx
     g_{eg}^2\left\langle\cos^2(2\eta)\right\rangle_{th}
     +
     \left(\frac{\hbar\omega S }{2}\right)^2\left\langle\sin^2(2\eta)\right\rangle_{th}
\end{align}
For a resonant cavity, the mixing between the 
cavity and optical transition takes its maximum value with $\eta = \pi/4$.
Furthermore, for a mixing 
angle of $\eta = \pi/4$, the  contribution to 
$G_{LU}$ from the internal non-adiabatic coupling is exactly equal to zero
and is expected to be small compared to the
reorganization energy $\hbar\omega S$.
However, we can expand this about $X=0$
at the crossing point, take the thermal average over the vibrational degrees of freedom, 
and use $2\lambda_{eg} = \hbar\Omega_{Rabi}$ to describe the coupling to between the 
cavity and the molecule.
Expanding the mixing angle terms
for small values of the nuclear displacement
\begin{align}
    \langle \hat w^2\rangle_{th}
& \approx
\left(\frac{\hbar\omega S}{2}\right)^2
\left(
1+\frac{(n_{th}+1) \left(4 g_{eg}^2-(\hbar \omega S)^2\right)}{2(\hbar \Omega_{Rabi})^2}
\right)
+\cdots
\end{align}
where we have summed over a thermal distribution of initial vibronic
states and evaluated the density of states at the
UP/LP energy gap. 

\begin{figure}
    \centering
\includegraphics[width=\columnwidth]{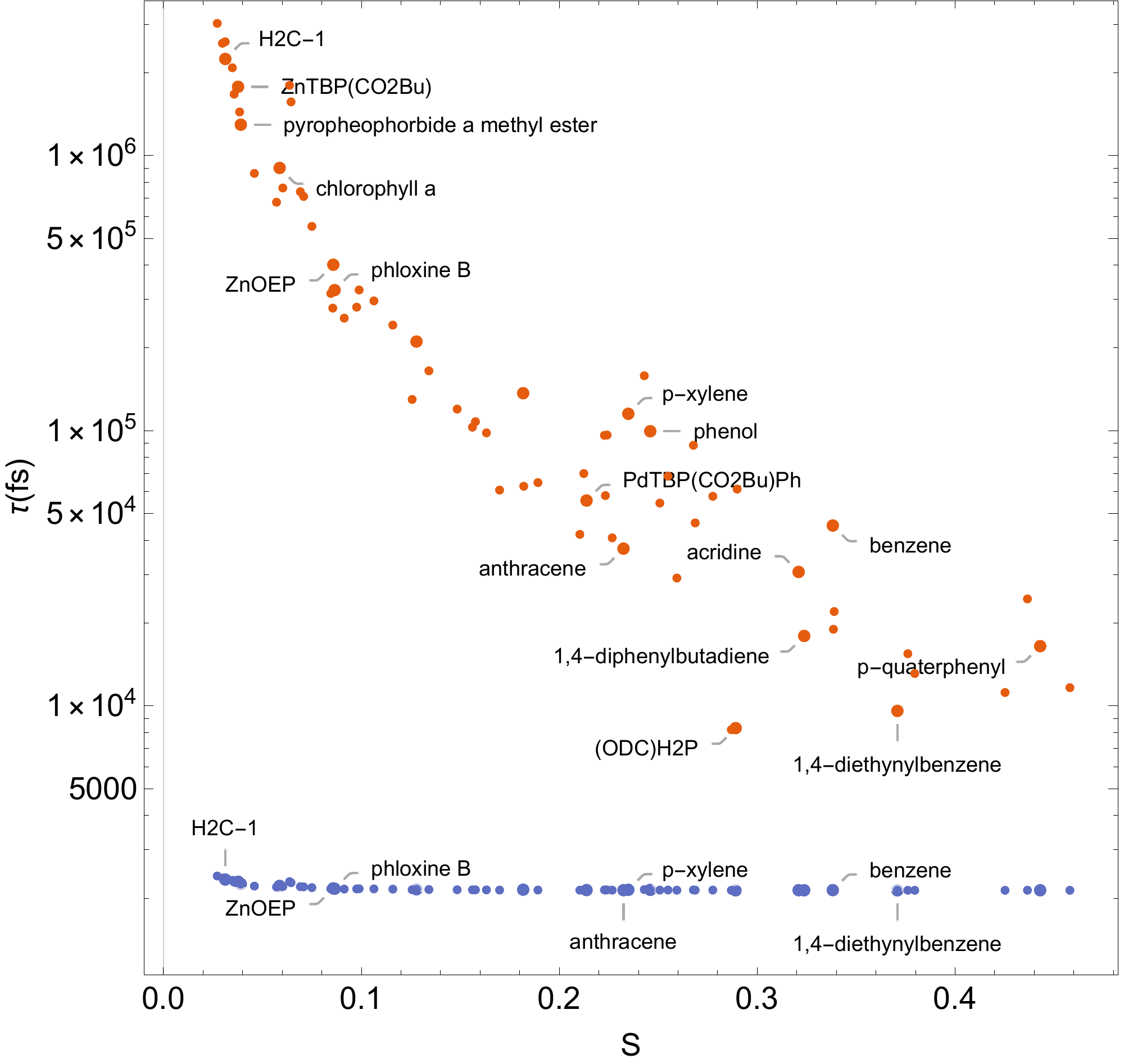}
    \caption{Golden Rule estimates of UP (red)
    and LP (blue)
    lifetimes for various molecular systems based upon their Huang-Rhys 
    factors, $S$ and vibronic frequencies.
    Data for selected molecules are labeled on
    each plot and a complete data table including
    references and structures for each system
    is provided as Supplemental Material.
    }
    \label{fig:plot}
\end{figure}
The treatment is valid for sufficiently 
small coupling  
such that $\hbar \omega S/2$ 
is small compared to the cavity line-width $\hbar \Gamma_{cav}$.
This means that the lifetime 
of the initial state UP is much longer than the lifetime  
for a cavity photon to decay 
into the continuum.   
This can be understood 
in terms of the period of a single Rabi oscillation
between the two states, which is on 
the order of $(S\omega)^{-1}$, 
and the time-scale for the decay of the LP state into
the continuum, which is $\Gamma_{cav}^{-1}$.  If the
LP state decays rapidly to the continuum, the system 
initially in UP
will decay immediately to the continuum once it passes through the LP
state  and have a vanishingly small likelihood 
of returning to the UP at the end of the first Rabi oscillation. Consequently, the lower limit of the
UP lifetime is set by the 
radiative decay of the cavity.

\subsection{Decay of the relaxed LP state}

We now consider the decay from the relaxed LP
state shown in Fig.~\ref{fig:poten}.  
In this scenario, we assume the 
LP can undergo vibronic relaxation from the
Franck-Condon point to the lower energy geometry
as depicted in Fig.~\ref{fig:poten}.  
As the system moves away from the Franck-Condon
point, the UP and LP states resolve back towards 
being purely cavity photon-like ($|g;1\rangle$)
and purely molecular
exciton-like ($|e;0\rangle$), respectively.

The analysis for the decay rate
proceeds as above, except that
the vibronic contribution is from the
shifted state  and that the 
the density of states contribution 
is evaluated at the UP-LP gap {\em at the relaxed
geometry}, i.e. $2\hbar\omega S$.
Only if $2\hbar\omega S$ is 
small compared to the natural line-width of the 
cavity $\hbar\Gamma_{cav}$ will this state be able to 
decay into the continuum. 
Consequently, systems with large reorganization
energies are more likely to form long-lived 
molecular excitons, held at their relaxed excited state
geometries.

\subsection{Molecular estimates}
We can use
the FGR expressions to provide estimates of the UP 
and LP lifetimes for a wide range of systems based upon their emission spectra.
For  organic conjugated systems, the vibronic fine
structure is dominated by C=C stretching modes 
of from 150-200 meV
with Huang-Rhys factors between 0.3 and 0.8. Table 1 in the Supplementary Information
gives
a summary of C=C frequencies and Huang-Rhys factors for various 
organic dye molecules and fluorophores obtained by analysing their fluorescence and absorption
spectra taken from literature sources. 
For this we assume a model cavity with $\hbar\Omega_{Rabi} = 400$\,meV
corresponding to a Rabi oscillation 
period of $\tau_{Rabi}=1.6$\,fs.
and a cavity photon lifetimes of 100\,fs 
corresponding to $\hbar\Gamma_{cav}=154$\,meV which are
typical in well-prepared molecular polariton systems.
Generally, conjugated molecular systems have
Huang-Rhys factors of $S \equiv 0.1 - 0.8$ and vibronic 
coupling is dominated by C=C stretching modes which 
are typically around $\hbar\omega = 200$\,meV.
Aside from the cavity couplings, all
parameters for our model can be directly extracted from 
linear absorption/emission spectra of a molecular system. 

We calculated reduced Huang-Rhys factors from spectral data freely available as part of the PhotochemCAD\texttrademark \ program and database for 
a variety of organic chromophores.  Out of the 339 compounds included 
in the PhotochemCAD\texttrademark \ database, we identified a subset of compounds with 1) absorption and emission spectra available and 2) an identifiable vibronic structure.
Using the peak position tool built into the PhotochemCAD\texttrademark \ program, we determined the vibrational energy, $\hbar \omega$, from the energy difference between adjacent vibronic peaks. 

Following the guidelines of de Jong \textit{et al.} 
\cite{C5CP02093J}
for determining $S$ from experimental spectra, we then converted each spectrum to an energy scale. Next we converted the absorbance and emission to transition moment squared, $W(E)$, using the relations $A(E) \propto EW(E)$ and $\phi(E) \propto E^{3}W(E)$, where $A(E)$ is the absorbance and $\phi(E)$ is the photon flux per unit of energy. We then calculated the barycenter of the baseline-corrected absorbance and emission spectra by evaluating
\begin{align}
    E_{bc} = \frac{\int W(E)E\,dE}{\int W(E)\,dE}
\end{align}
using Simpson's rule for each spectrum. Finally, for each compound we calculated the reduced Huang-Rhys parameter with the relation $\Delta E_{bc} = 2S$, where $\Delta E_{bc}$ is the difference between the barycenters of the absorption and emission spectra. 
A complete list of the couplings and sources is 
given in the SI.
From this data, we report estimated 
UP and LP decay times for a variety of common compounds in
Fig.~\ref{fig:plot}(a,b).

Fig~\ref{fig:plot}(a) , the model predicts a  broad range 
of UP lifetimes ranging from 100ps to less than 500 fs, largely 
determined by the variation in Huang-Rhys factors. Small Huang-Rhys
factors imply very weak vibronic coupling and small distortion 
of the molecule in its excited 
electronic state. 
Consequently, the 
system remains in the UP state near the avoided crossing region for long time.
The LP lifetime spans a much narrower range than the UP lifetime.
It is also surprisingly predicted to be shorter than the UP 
lifetime. 
This lifetime is entirely dominated by the 
fact that the relaxed LP is almost entirely excitonic-like 
($|e;0\rangle$) rather than photon-like ($|g;1\rangle$). 
Consequently, relaxation 
from this state via cavity emission 
is an activated process with  
$2\hbar\omega S \approx \hbar\Omega_{Rabi}$.


\section{Discussion}
The polariton lifetime dictates any eventual 
``polariton chemistry'' that might 
be induced or manipulated via coupling to the
cavity. Here, we consider the 
the lifetime for polaritons in resonant 
cavities, where we anticipate the
strongest coupling between the molecular
and photon degrees of freedom.  The long-lifetimes
of these states predicted by our model
is directly related to the 
fact that we prepare the system in a
light-matter eigenstate.  We also assume 
that the cavities themselves are of sufficiently
high-quality that strong-coupling can be achieved in the
sense that Rabi splitting between UP and LP branches 
is larger than the emission linewidths of the polaritons. 
These are reasonable assumptions that are consistent
with contemporary experimental conditions.

\begin{acknowledgments}
The work at the University of Houston was funded in
part by the  National Science Foundation (
CHE-1664971,   
CHE-1836080,  
DMR-1903785    
) and the Robert A. Welch Foundation (E-1337). 
GO acknowledges the support of the University of Houston
Honors College for a Summer Undergraduate
Research (SURF) fellowship. 
The work at Georgia Tech was funded by the National Science Foundation (DMR-1904293).

 
\end{acknowledgments}


\section*{Data Availability}

The data that support the findings of this study are available from the corresponding author upon reasonable request.


%

\clearpage
\section*{Supplementary Information}

\maketitle
\begin{widetext}
\section{Data Table}

    \hspace{-6cm}
    \begin{longtable}{cccccr}
    \textbf{Compound} &  \textbf{$\hbar \omega$ (eV)} &  \textbf{S (reduced)} &  \textbf{$\tau_{LP}$ (ps)} &  \textbf{$\tau_{UP}$ (ps)} & \textbf{References} \\ \hline \hline \hspace{0.5cm}
    \endfirsthead
    
    \textbf{Compound} &  \textbf{$\hbar \omega$ (eV)} &  \textbf{S (reduced)} &  \textbf{$\tau_{LP}$ (ps)} &  \textbf{$\tau_{UP}$ (ps)} & \textbf{References} \\ \hline \hline \hspace{0.25cm}
    \endhead
    \hline
    \multicolumn{6}{r}{{Continued on next page}} \\
    \endfoot
    \hline\endlastfoot
    \includegraphics[scale=0.1]{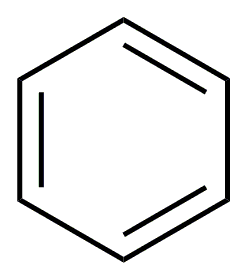}\\ benzene &    0.11 &         0.34 &              2.1 &             45 &    \citenum{ref11}, \citenum{ref50} \\
    \includegraphics[scale=0.1]{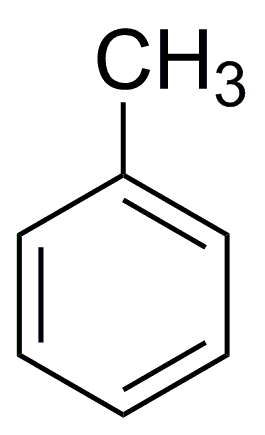}\\ toluene &  0.12 &         0.44 &              2.1 &             24 &                     \citenum{ref11} \\
    \includegraphics[scale=0.1]{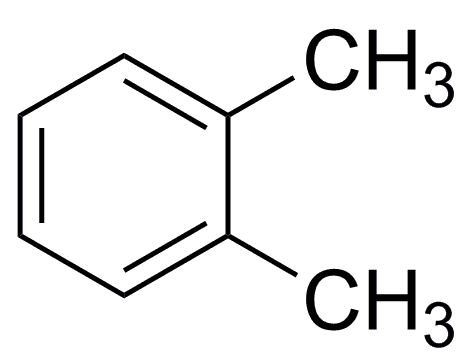}\\ o-xylene &            0.14 &         0.27 &              2.1 &             46 &    \citenum{ref54}, \citenum{ref55} \\
    \includegraphics[scale=0.1]{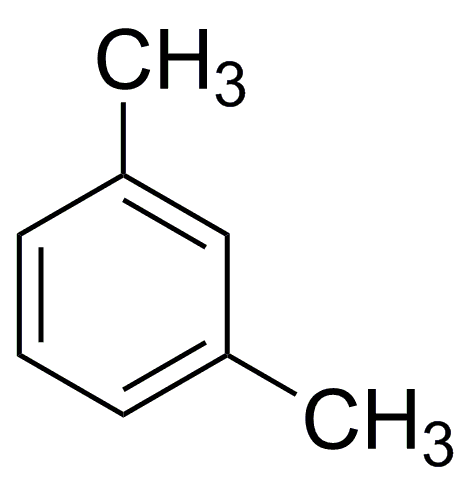}\\ m-oxylene &   0.14 &         0.25 &              2.1 &             55 &    \citenum{ref11}, \citenum{ref55} \\
    \includegraphics[scale=0.1]{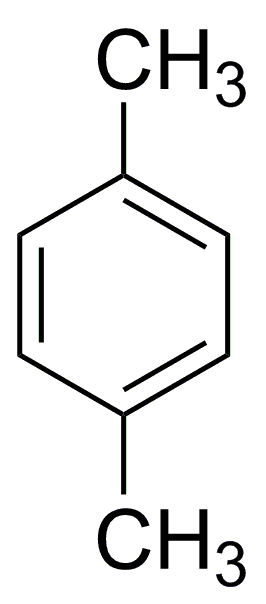}\\ p-xylene & 0.10 &         0.23 &              2.1 &            115 &    \citenum{ref11}, \citenum{ref55} \\
    \includegraphics[scale=0.1]{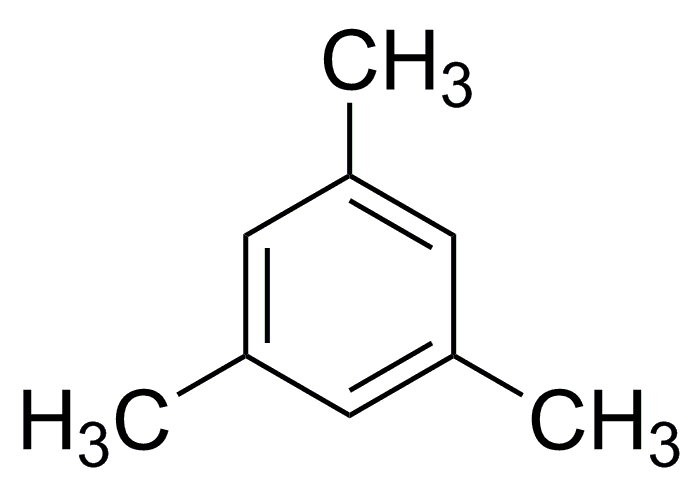}\\ mesitylene &             0.12 &         0.25 &              2.1 &             68 &    \citenum{ref55}, \citenum{ref56} \\
    \includegraphics[scale=0.1]{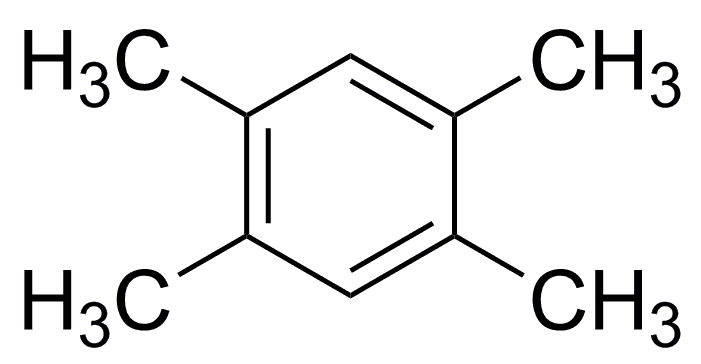}\\ durene &    0.15 &         0.22 &              2.1 &             58 &    \citenum{ref55}, \citenum{ref56} \\
    \includegraphics[scale=0.1]{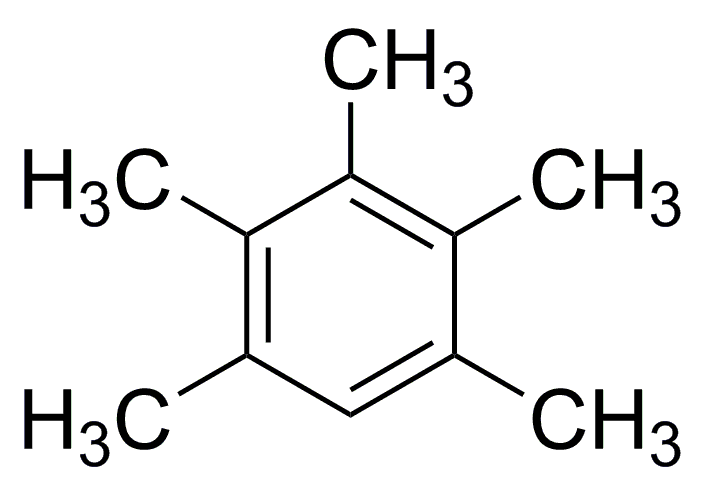}\\ pentamethylbenzene &        0.08 &         0.24 &              2.1 &            158 &    \citenum{ref55}, \citenum{ref56} \\
    \includegraphics[scale=0.1]{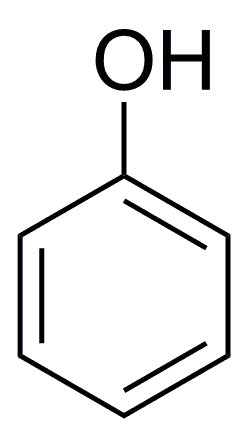}\\phenol &       0.10 &         0.25 &              2.1 &             99 &    \citenum{ref11}, \citenum{ref62} \\
    \includegraphics[scale=0.1]{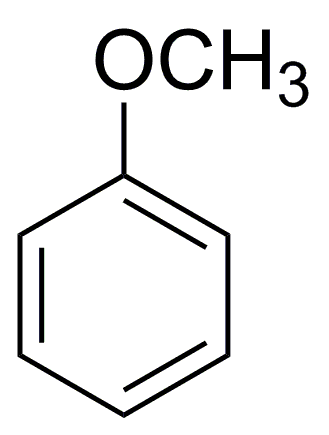}\\anisole &    0.12 &         0.22 &              2.1 &             96 &                     \citenum{ref64} \\
    \includegraphics[scale=0.1]{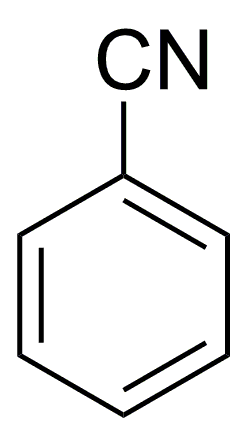}\\benzonitrile &       0.12 &         0.22 &              2.1 &             96 &    \citenum{ref64}, \citenum{ref66} \\
    \includegraphics[scale=0.1]{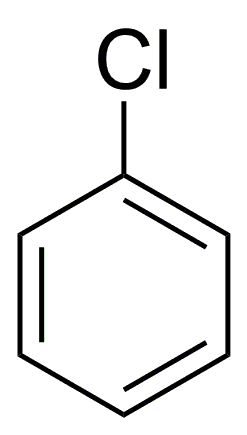}\\chlorobenzene &    0.12 &         0.28 &              2.1 &             58 &    \citenum{ref85}, \citenum{ref86} \\
    \includegraphics[scale=0.1]{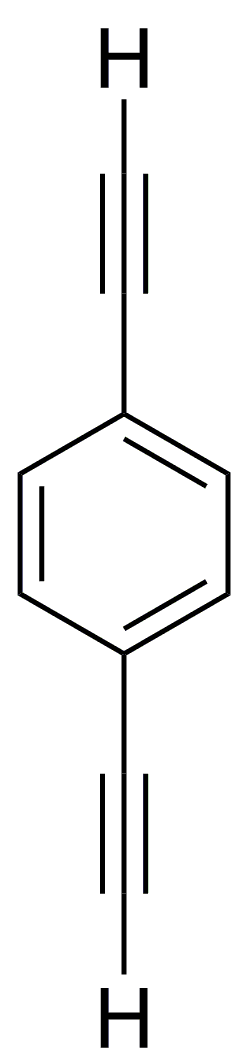}\\1,4-diethynylbenzene &            0.23 &         0.37 &              2.1 &              9.5 &                    \citenum{ref100} \\
    \includegraphics[scale=0.1]{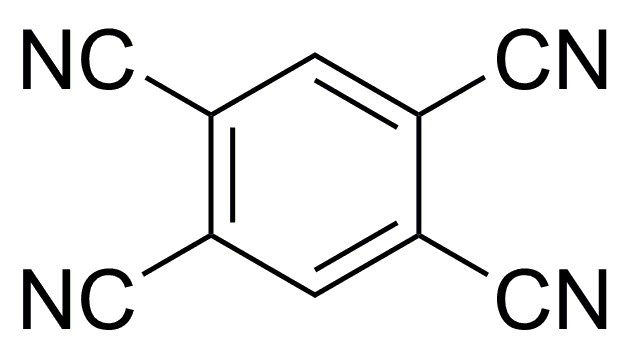}\\1,2,4,5-tetracyanobenzene &     0.15 &         0.16 &              2.1 &            107 &                    \citenum{ref129} \\
    \includegraphics[scale=0.1]{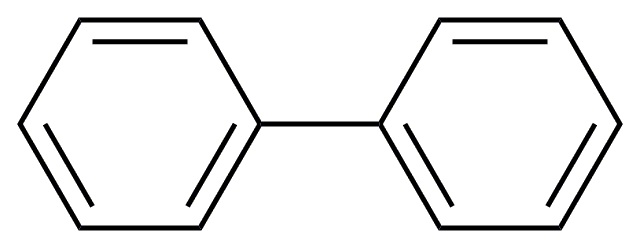} \\ biphenyl &  0.16 &         0.61 &              2.1 &              7.3 &                     \citenum{ref11} \\
   \includegraphics[scale=0.1]{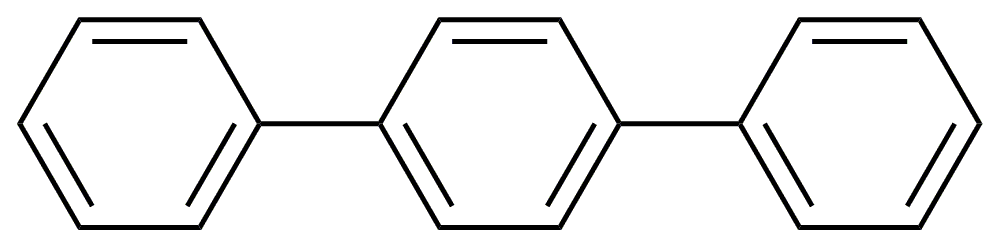}\\  p-terphenyl &      0.16 &         0.46 &              2.1 &             11.6 &                     \citenum{ref11} \\
   \includegraphics[scale=0.1]{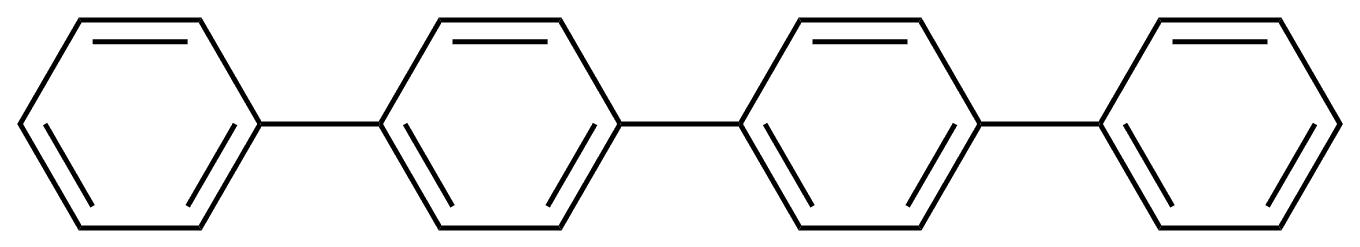} \\ p-quaterphenyl &     0.14 &         0.44 &              2.1 &             16.5 &                     \citenum{ref11} \\
    \includegraphics[scale=0.1]{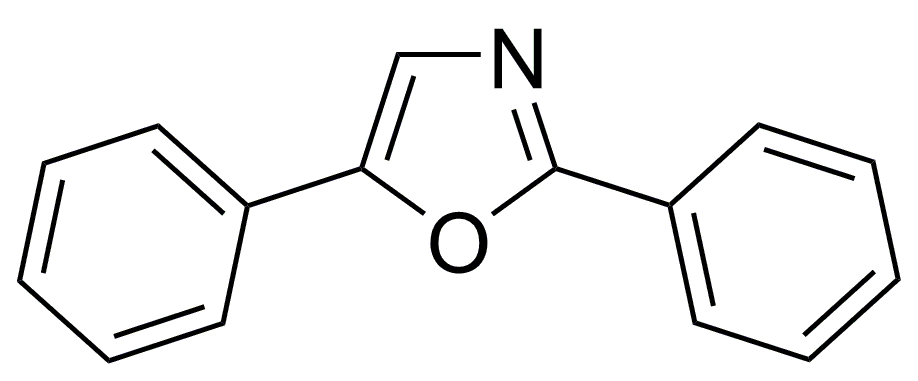}\\ 2,5-diphenyloxazole [PPO] &      0.19 &         0.38 &              2.1 &             13.0 &                     \citenum{ref11} \\
    \includegraphics[scale=0.8]{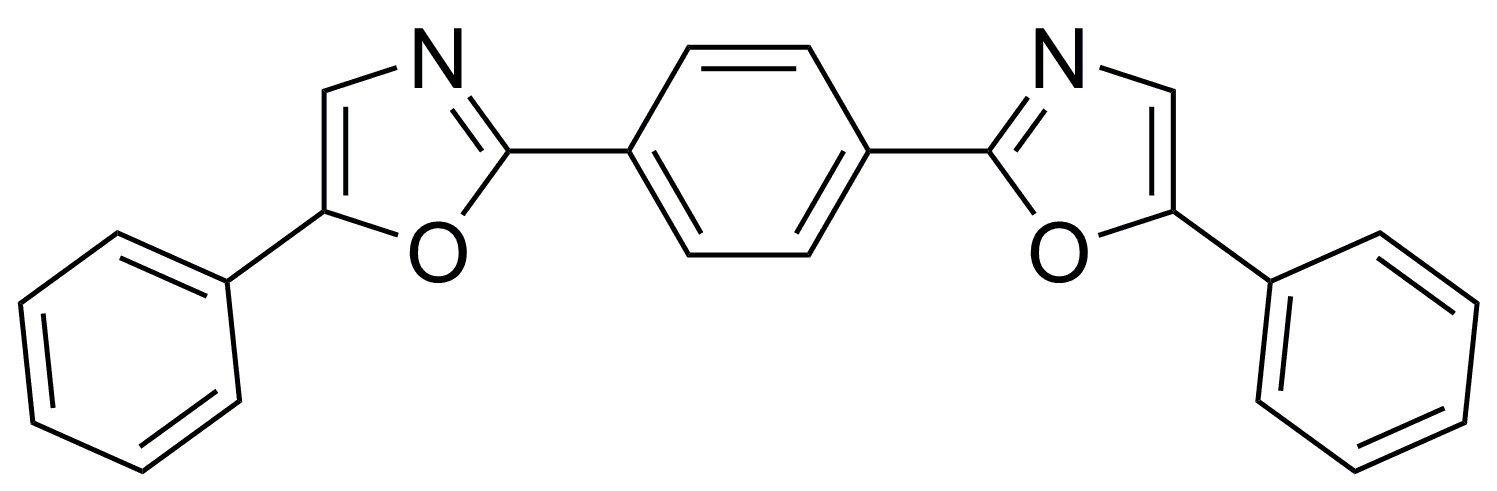} \\ 1,4-Bis(5-phenyl-2-oxazolyl)benzene, [POPOP] &      0.18 &         0.26 &              2.1 &             29.1 &                     \citenum{ref11} \\
    \includegraphics[scale=0.1]{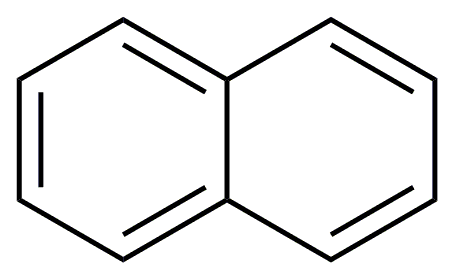}\\ naphthalene &   0.17 &         0.38 &              2.1 &             15454 &                     \citenum{ref11} \\
    \includegraphics[scale=0.1]{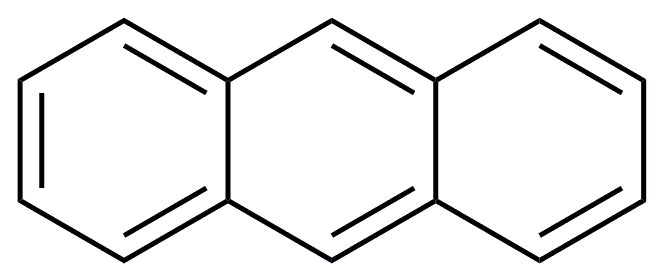}\\  anthracene &    0.18 &         0.23 &              2.1 &             37.2 &                     \citenum{ref11} \\
    \includegraphics[scale=0.1]{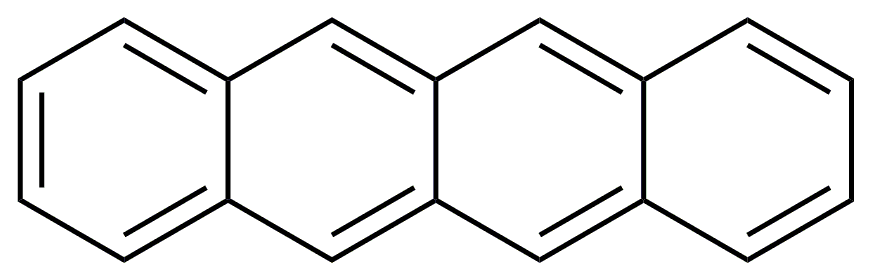}\\ tetracene &   0.18 &         0.18 &              2.1 &             62.7 &   \citenum{ref11}, \citenum{ref155} \\
    \includegraphics[scale=0.1]{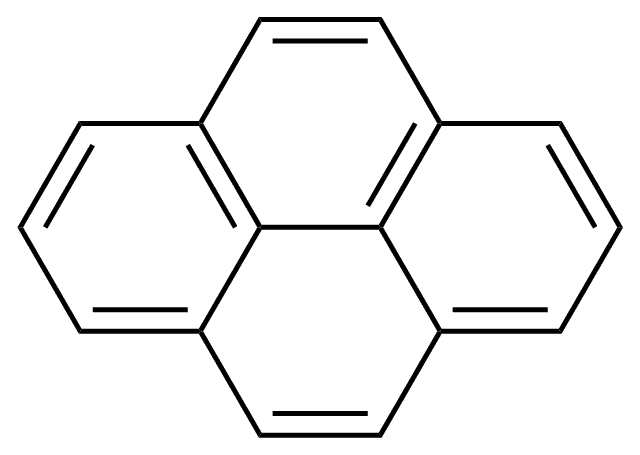} \\ pyrene &      0.17 &         0.34 &              2.1 &             19.0 &                     \citenum{ref11} \\
    \includegraphics[scale=0.1]{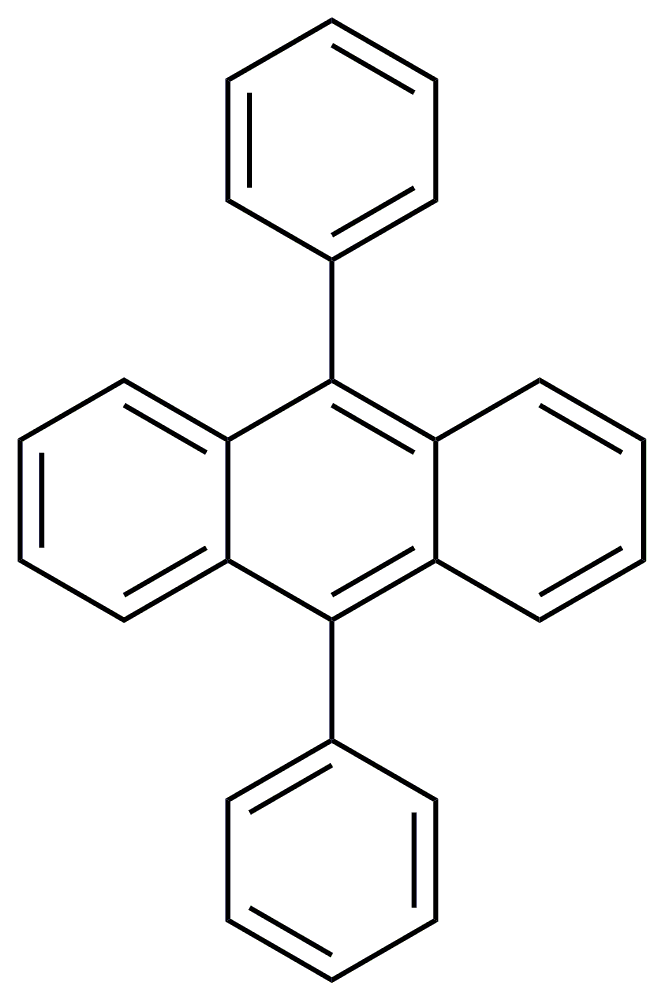} \\  9,10-diphenylanthracene &    0.18 &         0.23 &              2.1 &             40.7 &                     \citenum{ref11} \\
    \includegraphics[scale=0.1]{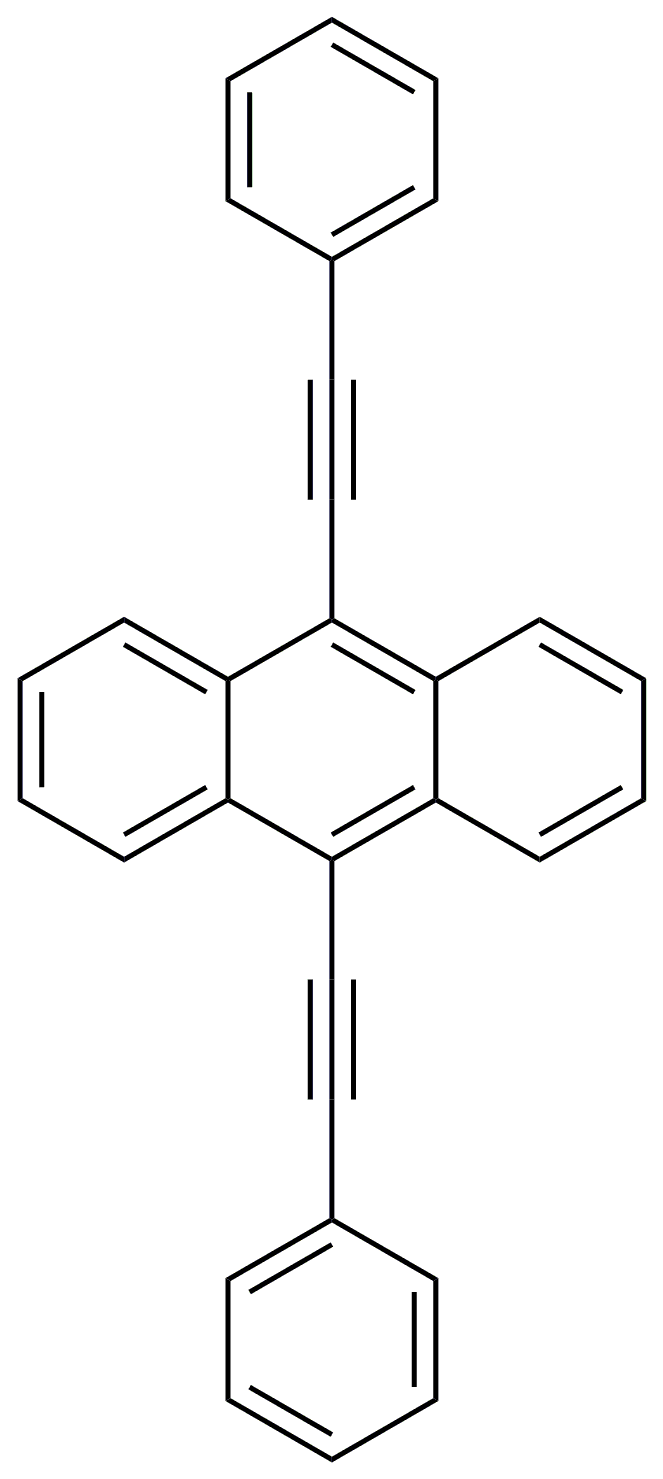}\\ 9,10-Bis(phenylethynyl)anthracene &     0.12 &         0.18 &              2.1 &            136 &                     \citenum{ref11} \\
    \includegraphics[scale=0.1]{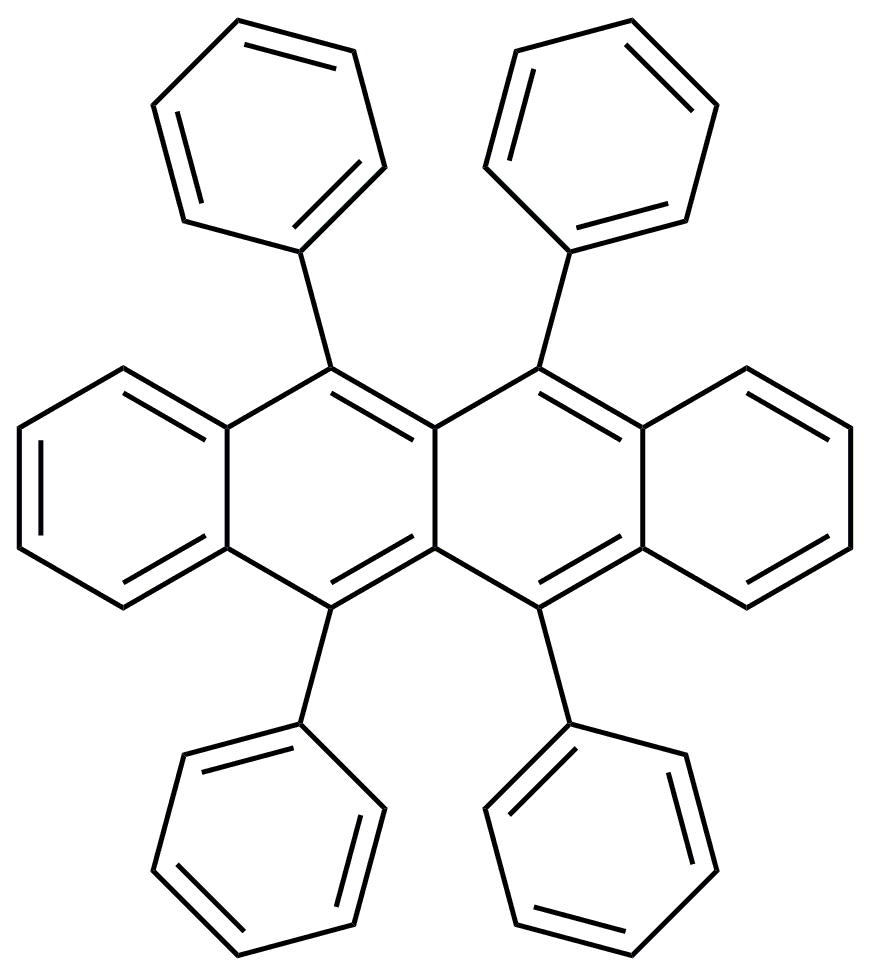}\\ rubrene &     0.17 &         0.19 &              2.1 &             64.7 &   \citenum{ref11}, \citenum{ref162} \\
    \includegraphics[scale=0.1]{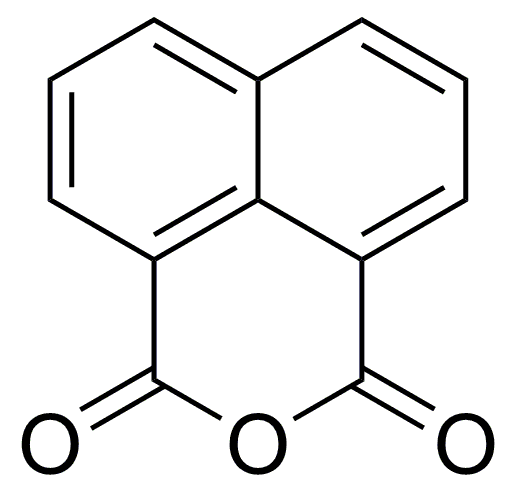} \\ 1,8-naphthalic anhydride &       0.11 &         0.29 &              2.1 &             61.3 &                    \citenum{ref164} \\
    \includegraphics[scale=0.1]{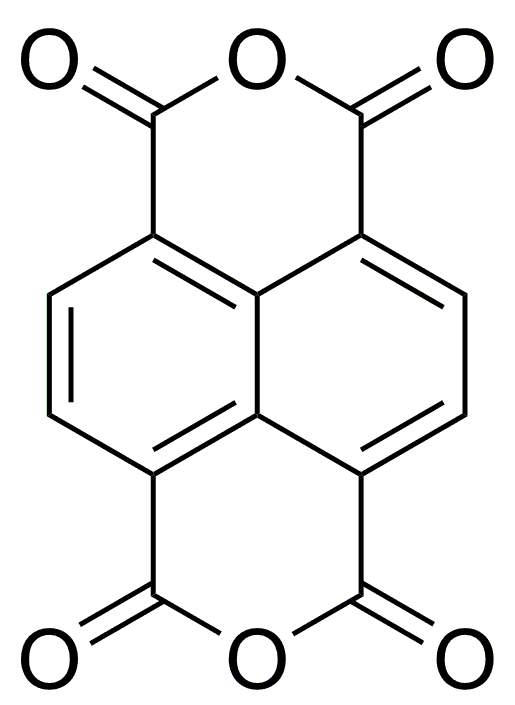}\\ 1,4,5,8-naphthalenetetracarboxylic dianhydride &       0.19 &         0.21 &              2.1 &             42.0 &                    \citenum{ref173} \\
    \includegraphics[scale=0.1]{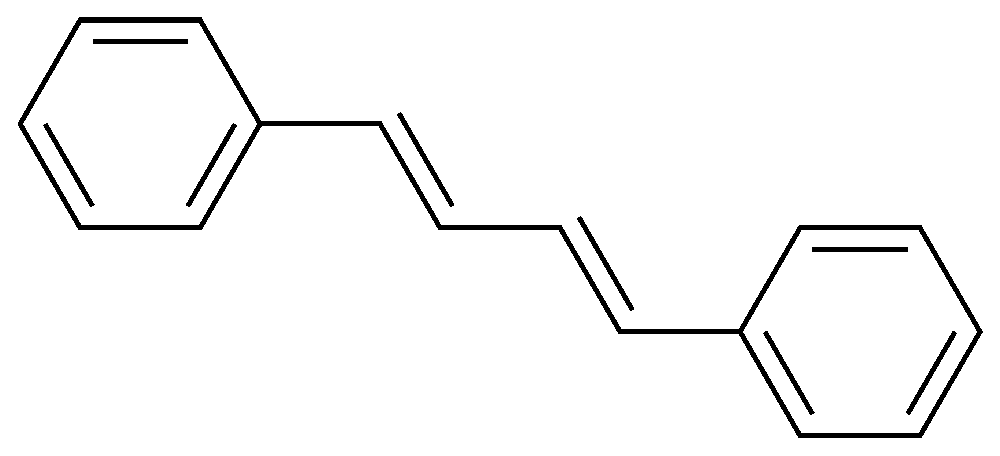}\\ 1,4-diphenylbutadiene &    0.19 &         0.32 &              2.1 &             17.9 &  \citenum{ref186}, \citenum{ref187} \\
    \includegraphics[scale=0.1]{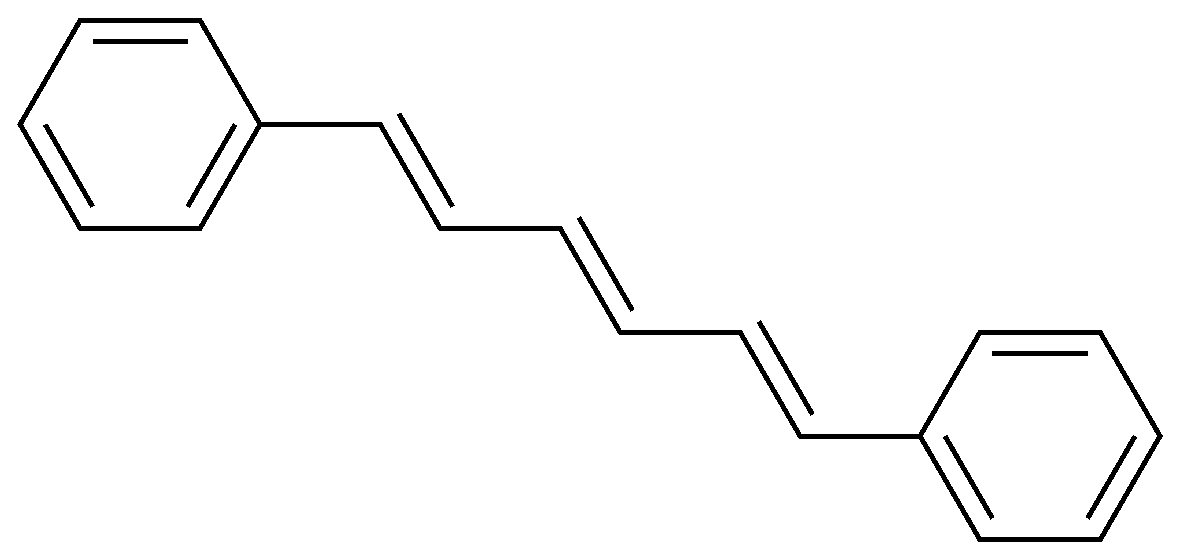}\\  1,6-diphenylhexatriene &        0.18 &         0.43 &              2.1 &             11.2 &                    \citenum{ref187} \\
    \includegraphics[scale=0.1]{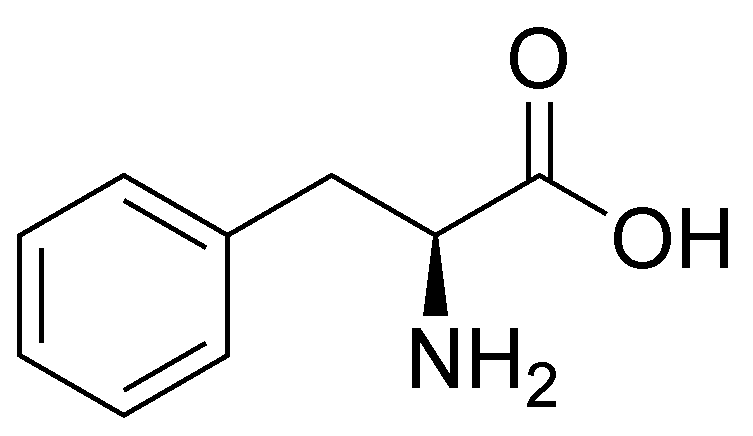} \\  L-phenylalanine &   0.10 &         0.27 &              2.1 &             88.4 &   \citenum{ref38}, \citenum{ref263} \\
    \includegraphics[scale=0.1]{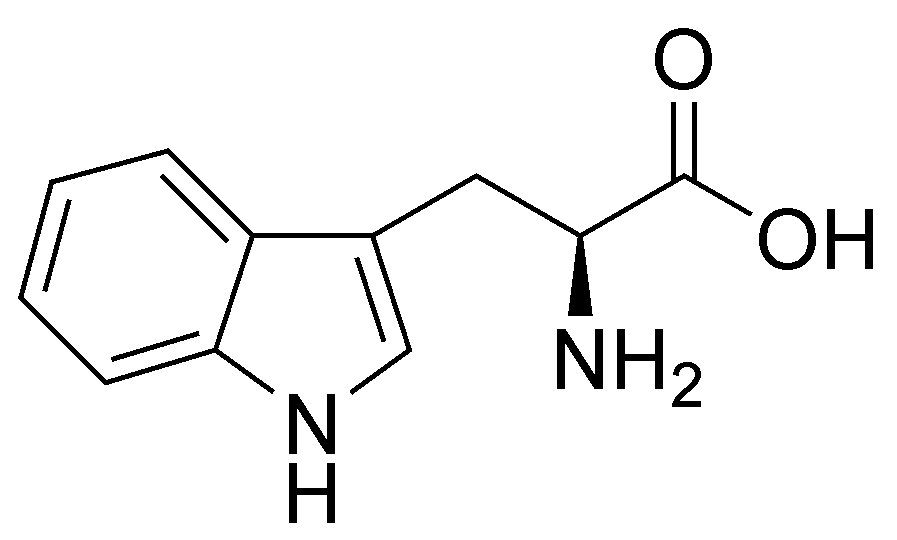}\\ L-tryptophan &  0.15 &         0.58 &              2.1 &              9.31 &   \citenum{ref38}, \citenum{ref263} \\
    \includegraphics[scale=.9]{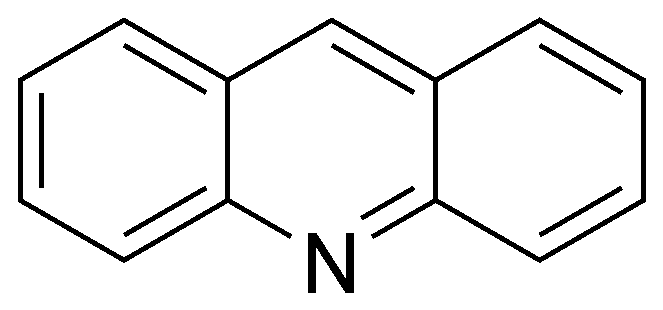} \\  acridine &    0.14 &         0.32 &              2.1 &             30.6 &  \citenum{ref376}, \citenum{ref377} \\
    \includegraphics[scale=0.1]{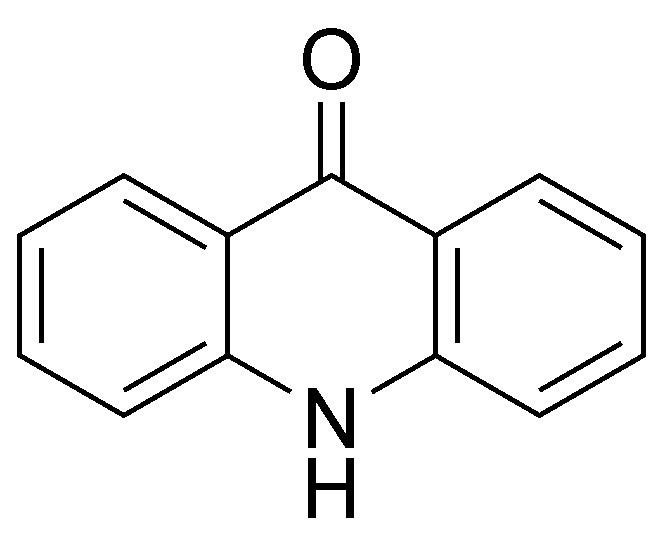}\\ acridone &    0.16 &         0.34 &              2.1 &             22.0 &   \citenum{ref50}, \citenum{ref386} \\
    \includegraphics[scale=0.1]{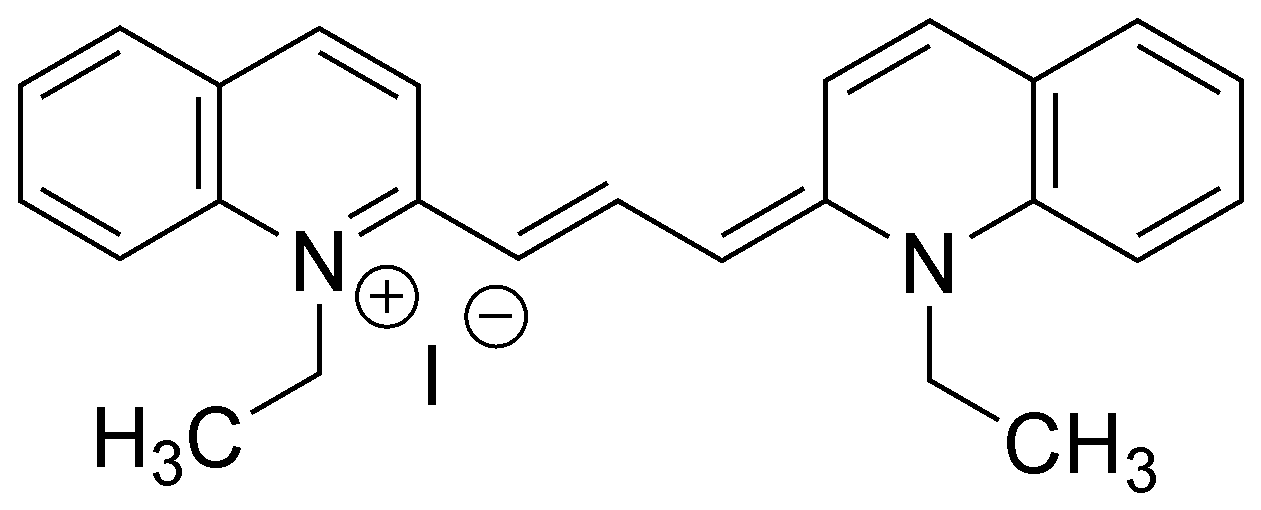}\\ 1,1'-diethyl-2,2'-carbocyanine iodide &       0.15 &         0.10 &              2.2 &            280 &                    \citenum{ref482} \\
    \includegraphics[scale=0.1]{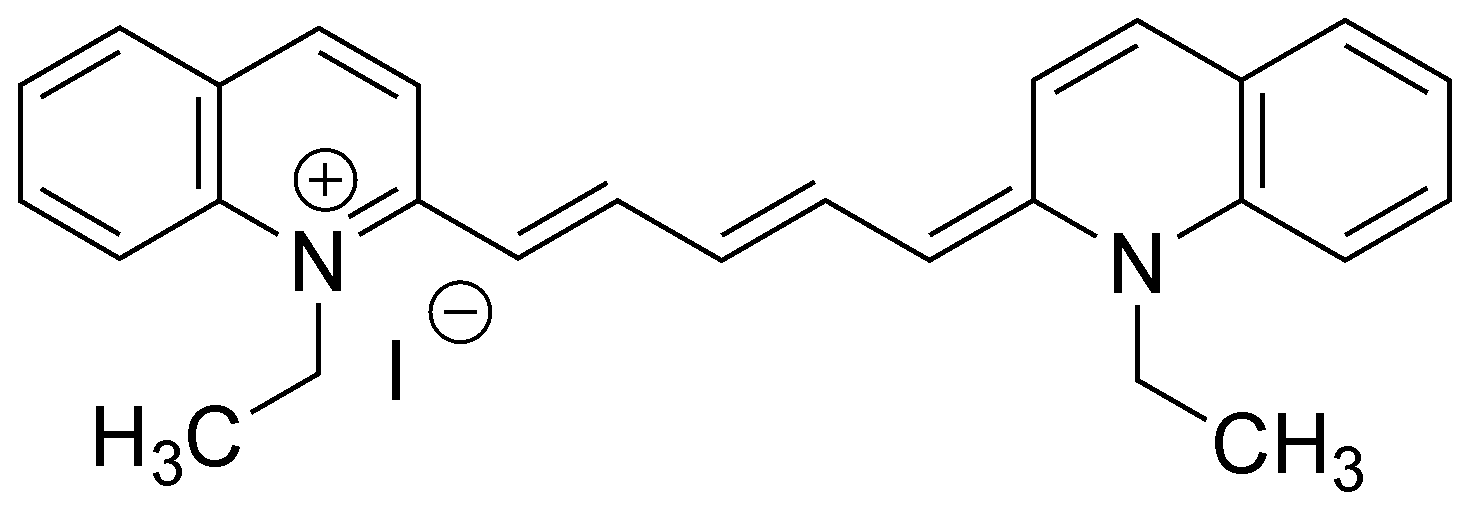}\\  1,1'-diethyl-2,2'-dicarbocyanine iodide &        0.15 &         0.06 &              2.2 &            761 &                    \citenum{ref485} \\
    \includegraphics[scale=0.1]{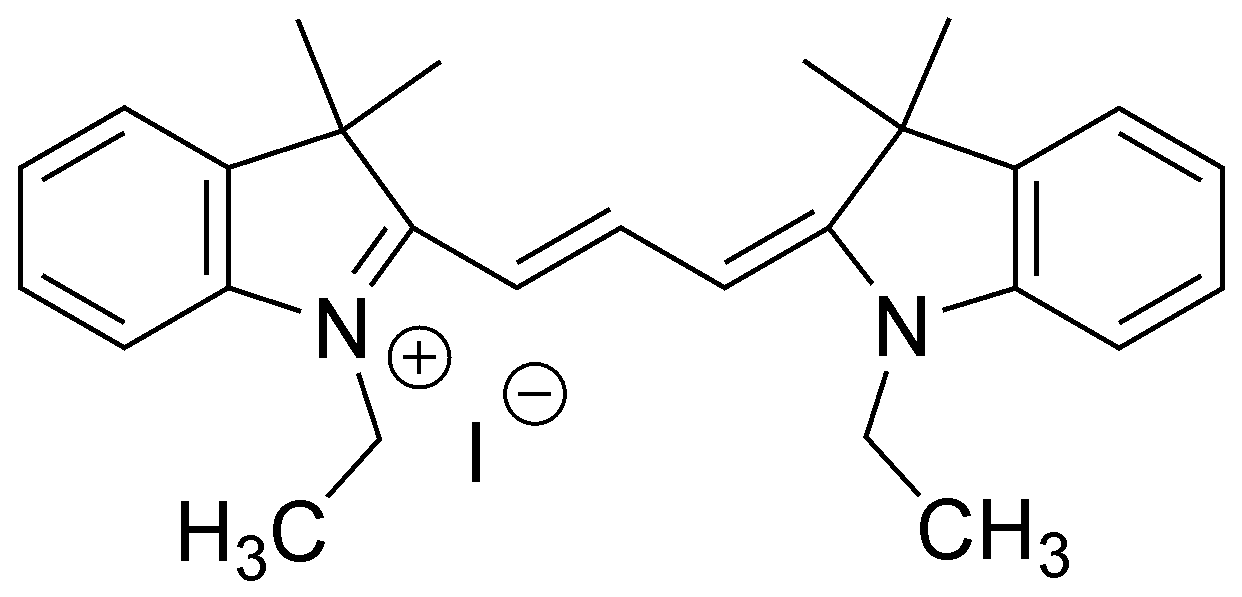}\\ 1,1'-diethyl-3,3,3',3'-tetramethylindocarbocyanine iodide &      0.14 &         0.13 &              2.1 &            211 &  \citenum{ref487}, \citenum{ref488} \\
    \includegraphics[scale=0.1]{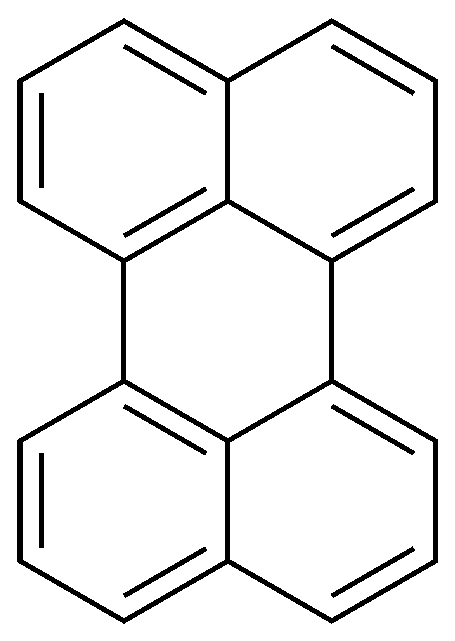} \\ perylene &       0.19 &         0.17 &              2.1 &             61.0 &                     \citenum{ref11} \\
    \includegraphics[scale=0.1]{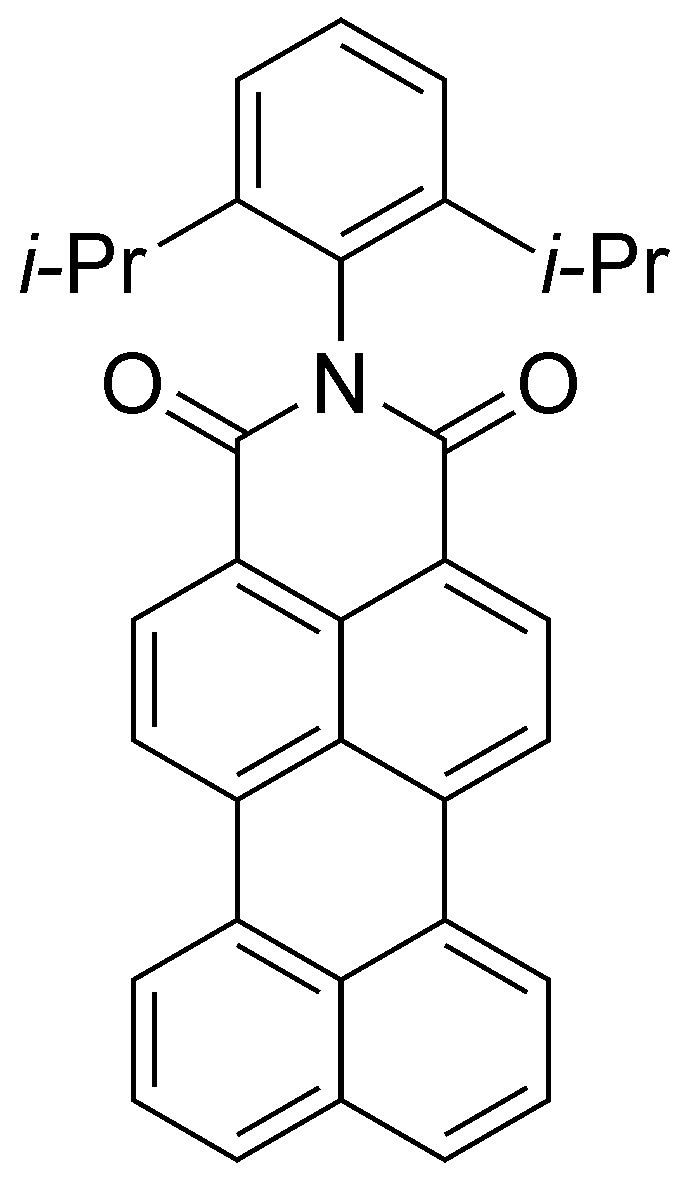}\\ perylene, PMI &    0.14 &         0.21 &              2.1 &             70.0 &                    \citenum{ref533} \\
    \includegraphics[scale=0.1]{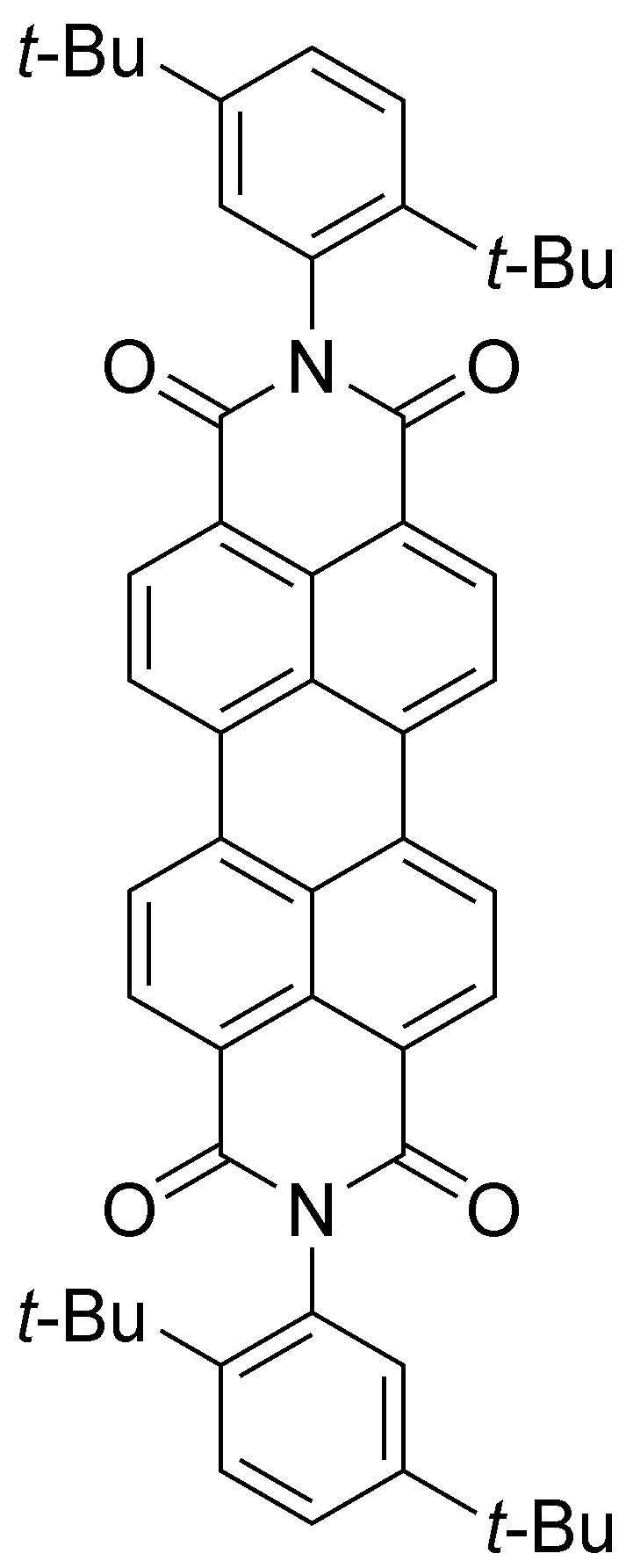}\\ perylene, PDI &    0.18 &         0.13 &              2.1 &            130 &  \citenum{ref535}, \citenum{ref536} \\
    \includegraphics[scale=0.1]{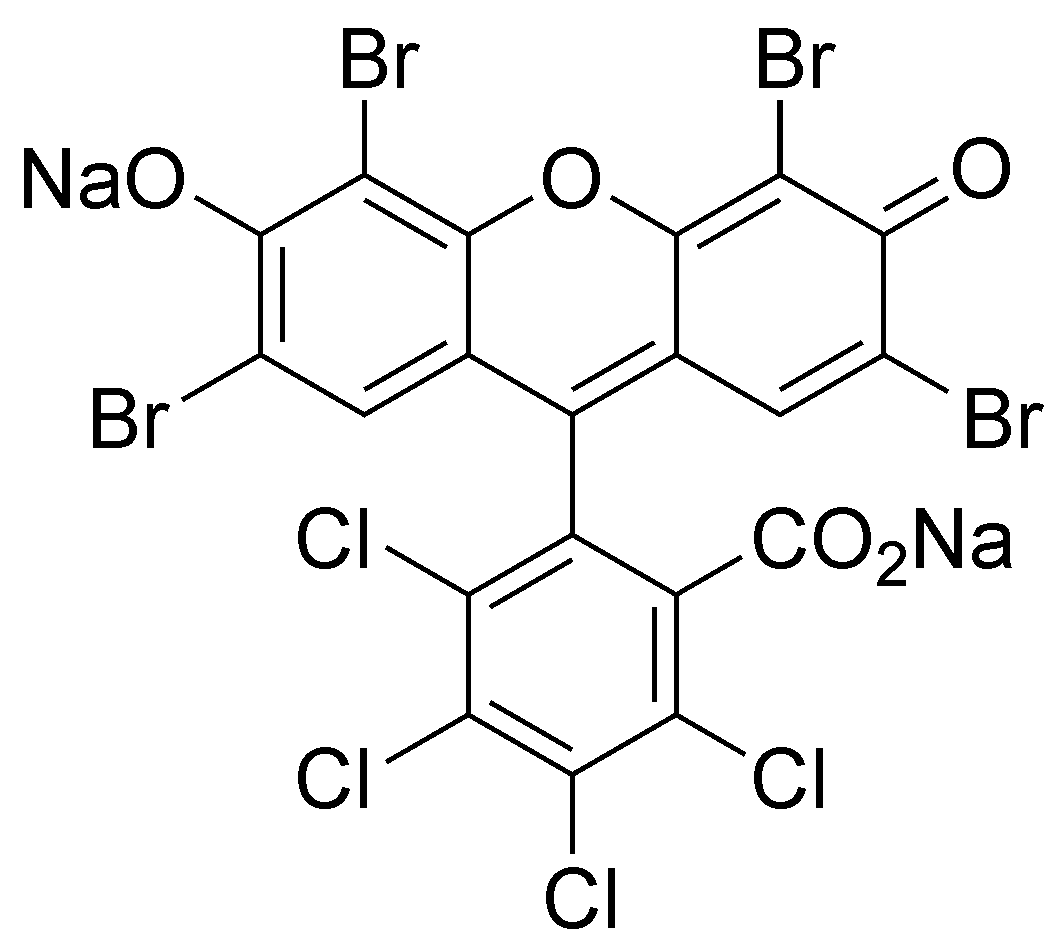}\\ phloxine B &    0.16 &         0.09 &              2.2 &            324 &                    \citenum{ref550} \\
    \includegraphics[scale=0.1]{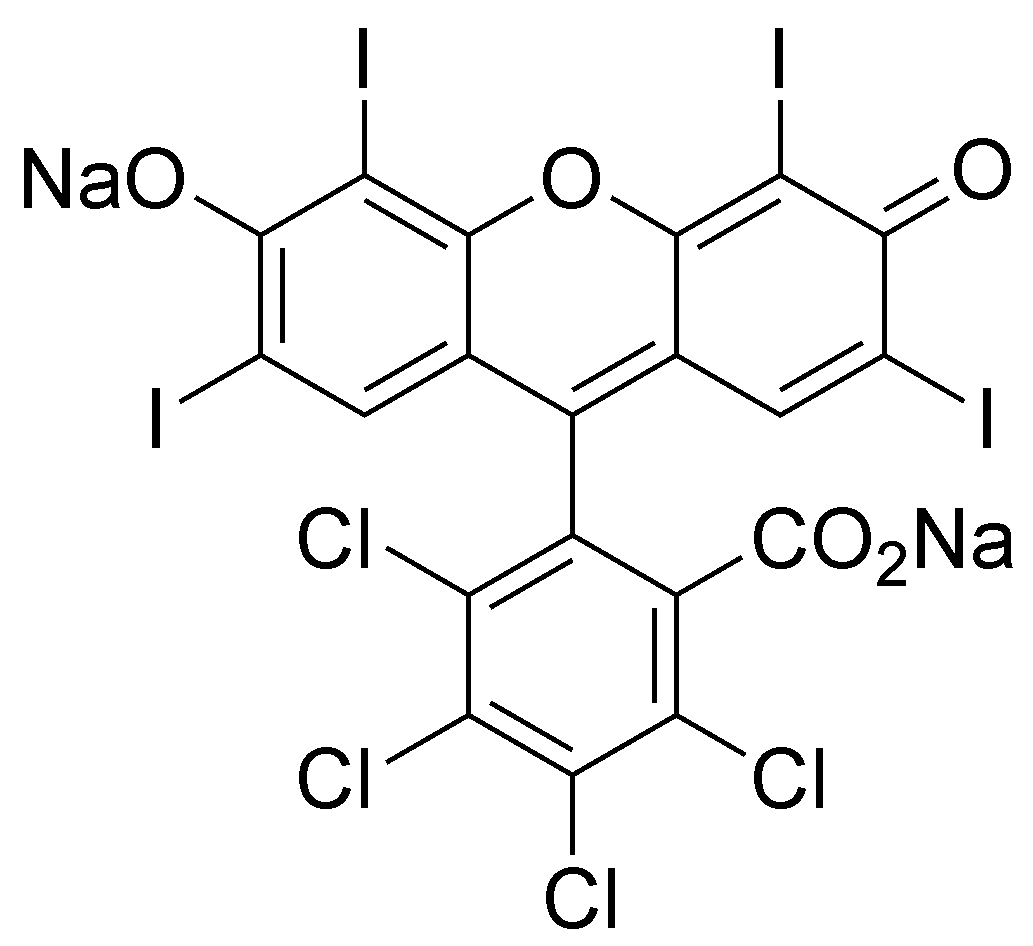}\\ rose bengal &      0.17 &         0.08 &              2.2 &            315 &                    \citenum{ref538} \\
    \includegraphics[scale=0.1]{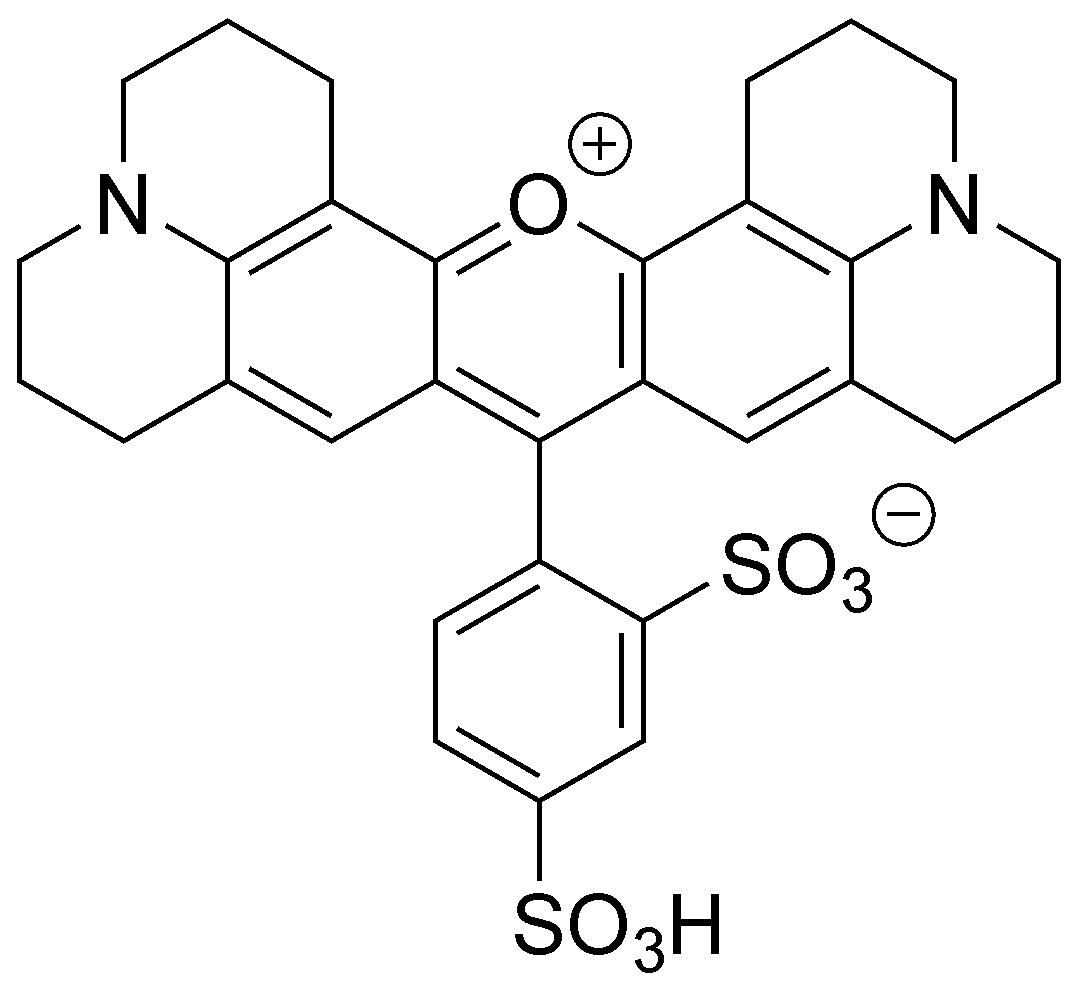}\\ sulforhodamine 101 &       0.14 &         0.08 &              2.2 &            552 &  \citenum{ref573}, \citenum{ref574} \\
    \includegraphics[scale=0.1]{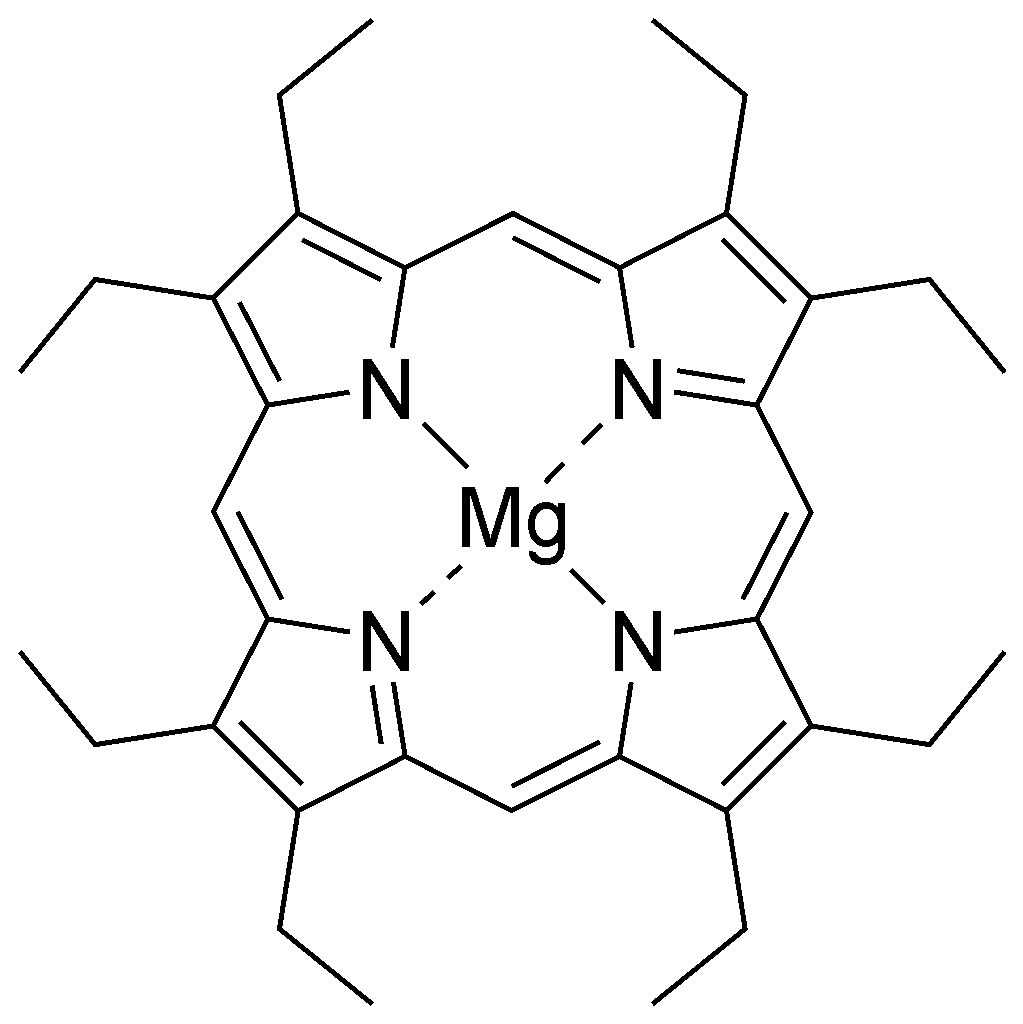} \\  MgOEP &      0.14 &         0.10 &              2.2 &            324 &                    \citenum{ref620} \\
    \includegraphics[scale=0.1]{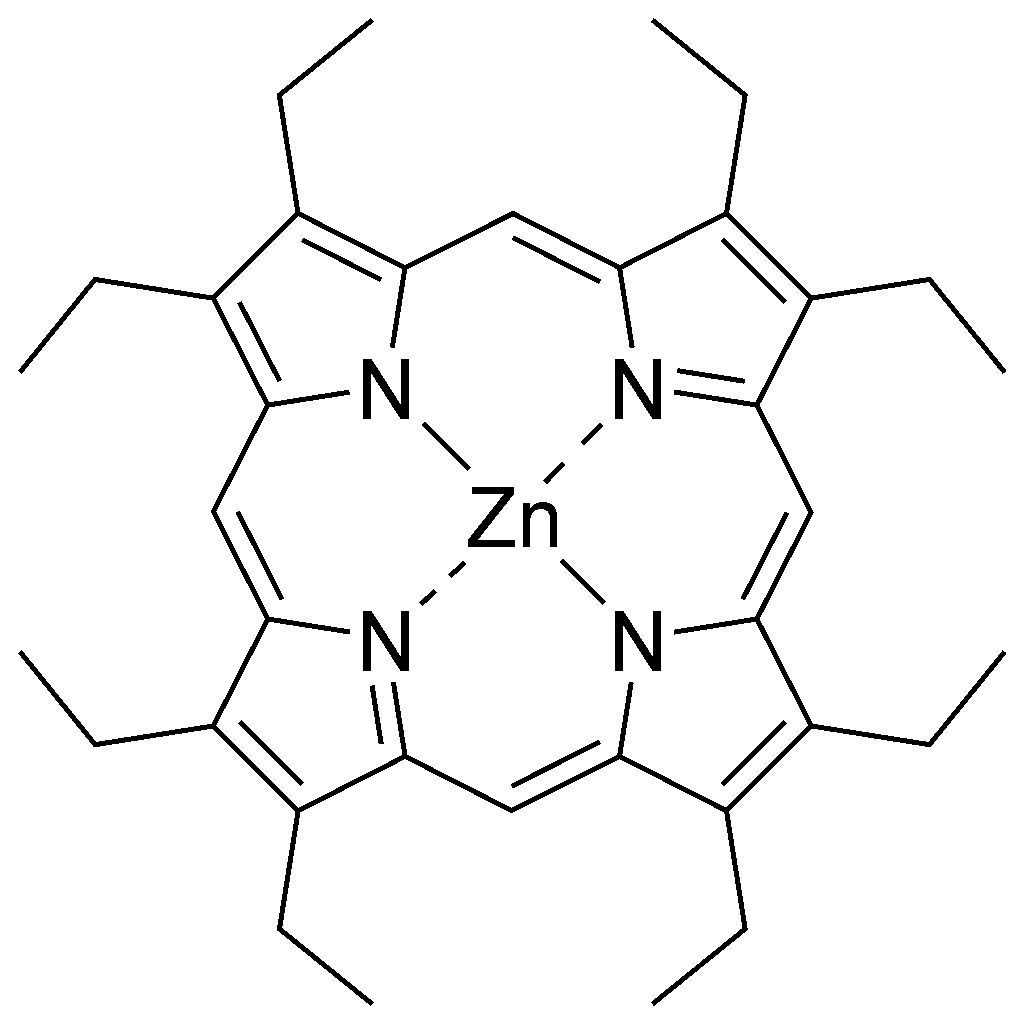}\\ ZnOEP &        0.15 &         0.09 &              2.2 &            401 &  \citenum{ref613}, \citenum{ref623} \\
    \includegraphics[scale=0.1]{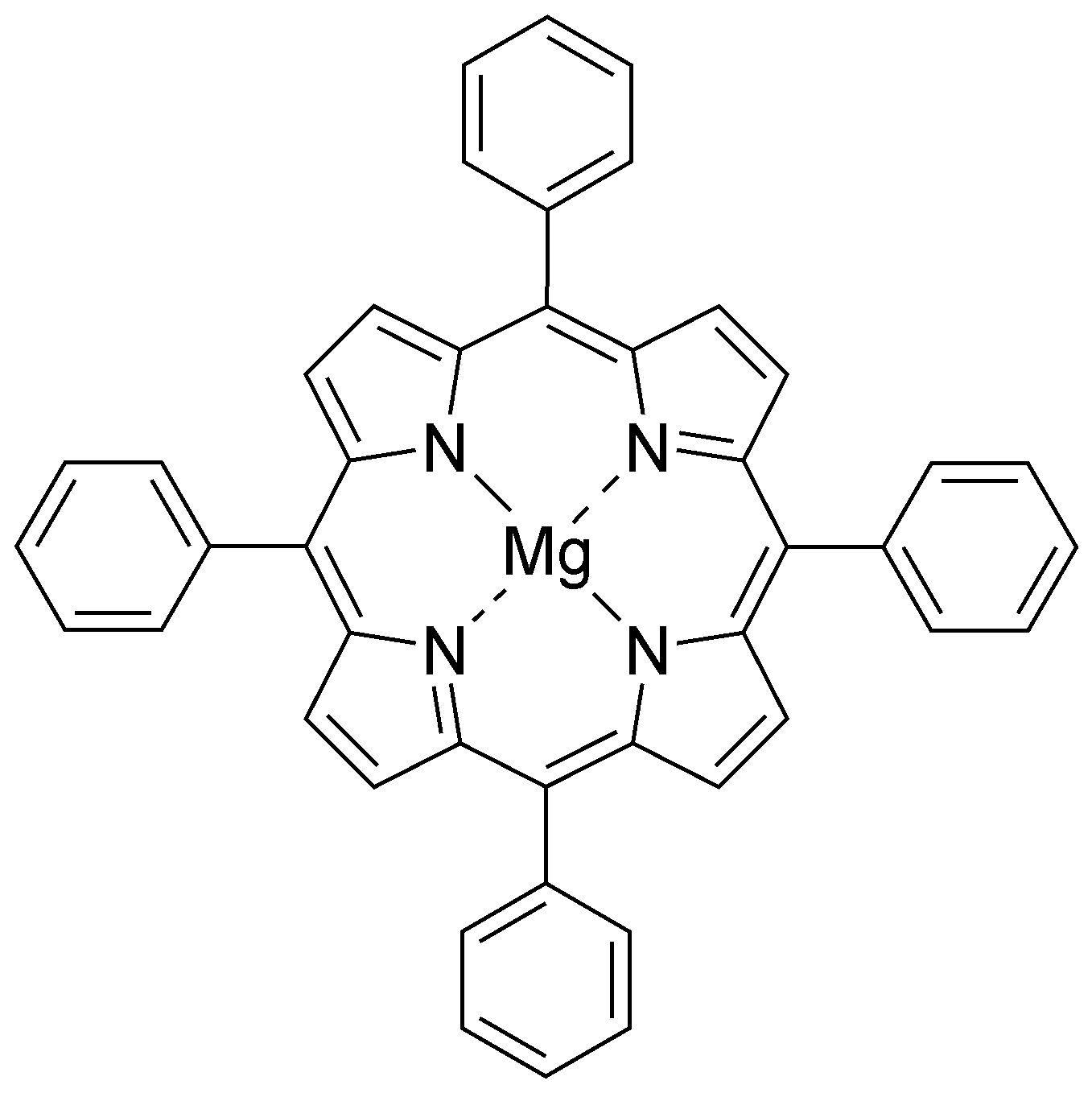}\\ MgTPP &        0.14 &         0.12 &              2.2 &            242 &  \citenum{ref632}, \citenum{ref633} \\
    \includegraphics[scale=0.1]{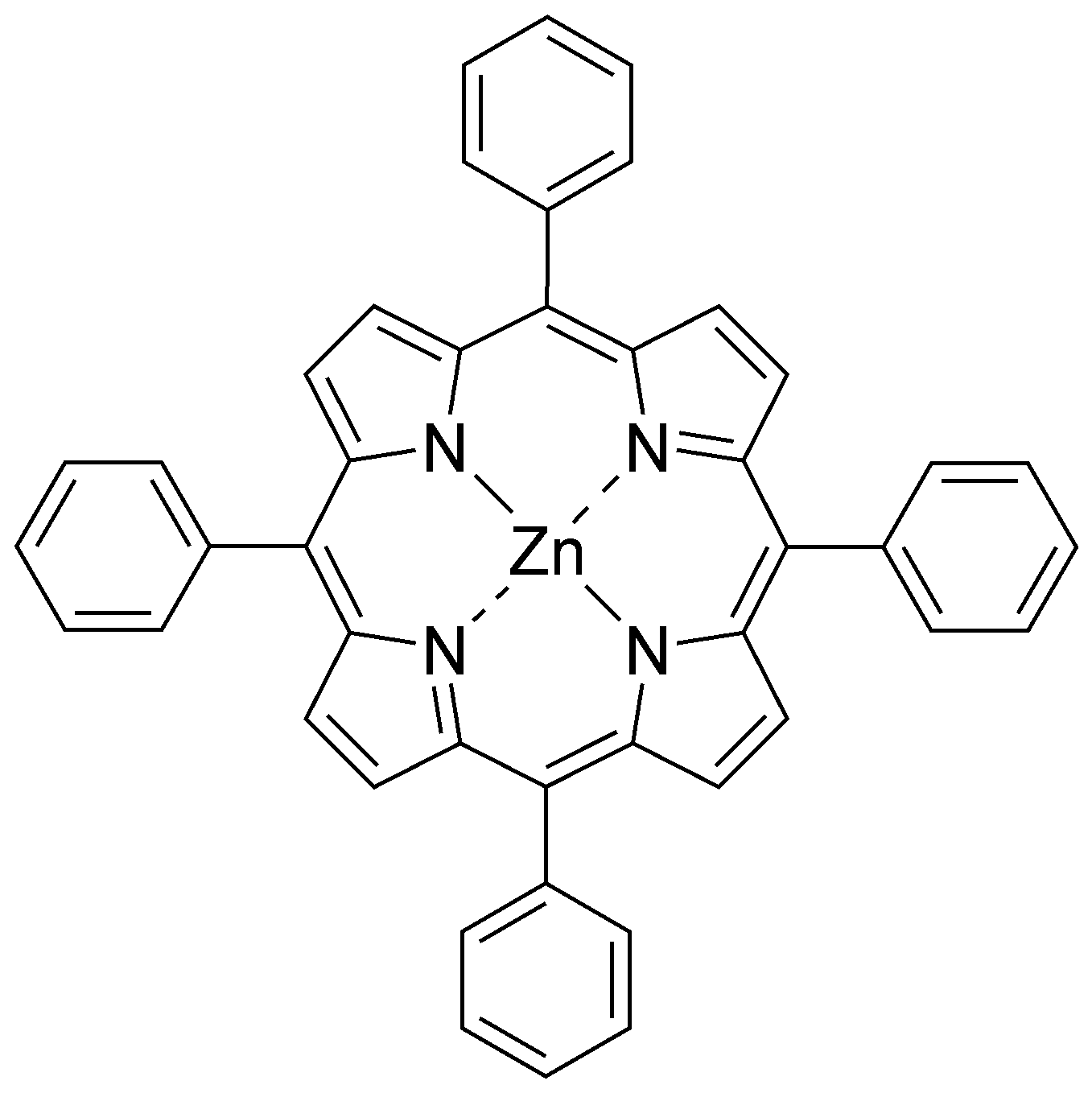}\\ ZnTPP &    0.16 &         0.15 &              2.1 &            120 &  \citenum{ref629}, \citenum{ref633} \\
    \includegraphics[scale=0.1]{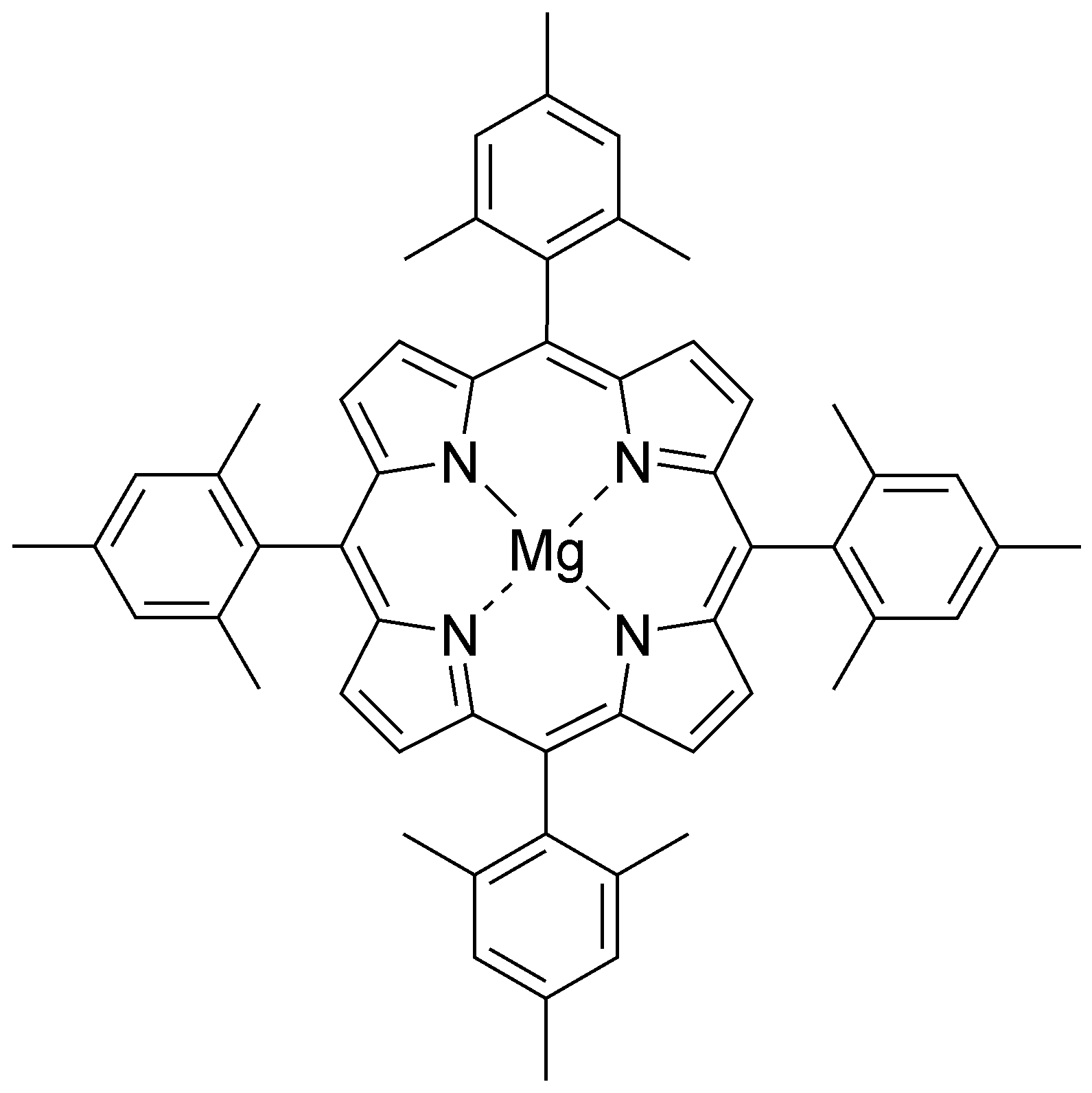}\\ MgTMP &       0.15 &         0.13 &              2.1 &            165 &  \citenum{ref638}, \citenum{ref639} \\
    \includegraphics[scale=0.1]{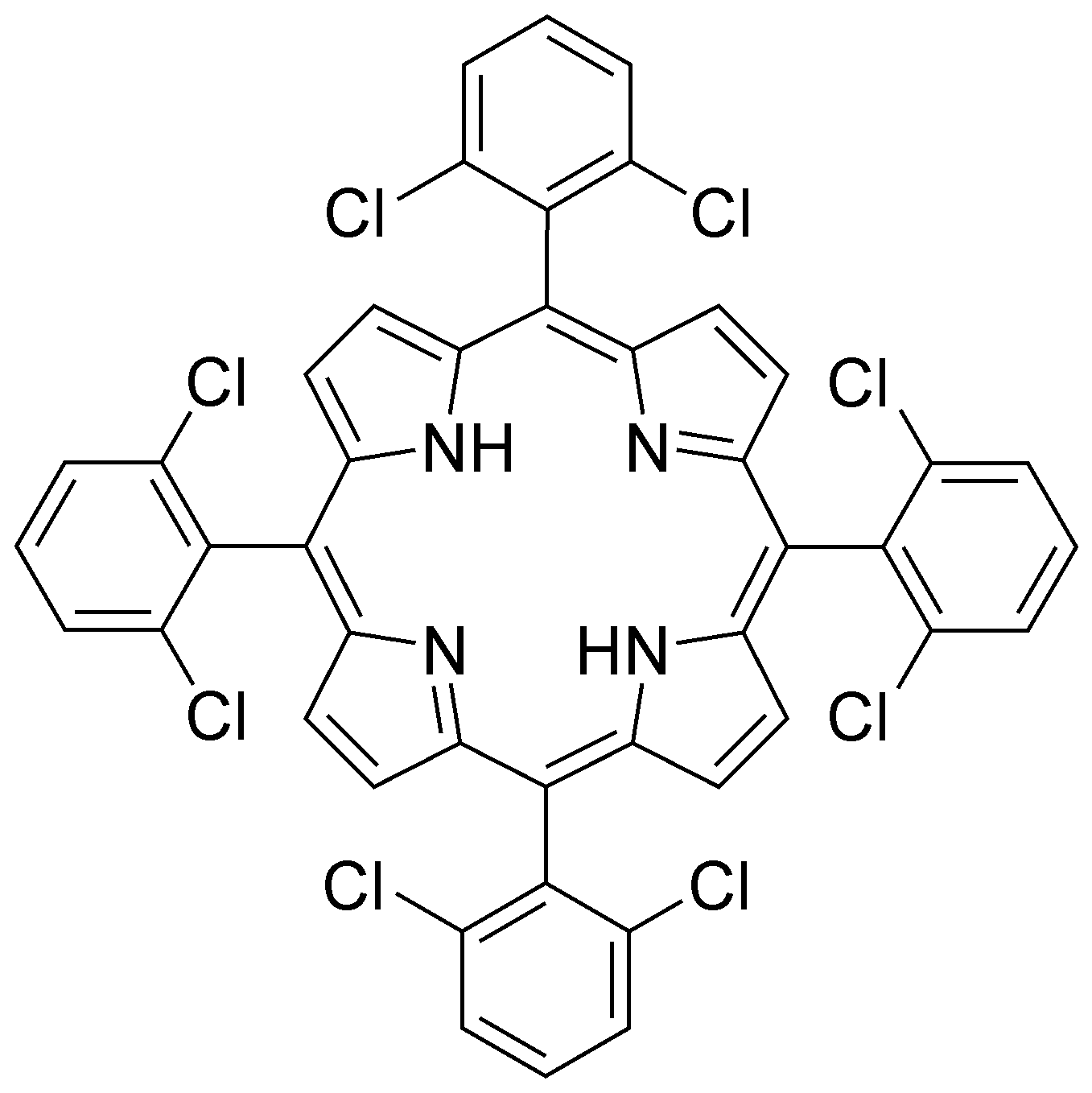}\\ (ODC)H2P &     0.31 &         0.29 &              2.1 &              8.29 &  \citenum{ref627}, \citenum{ref638} \\
    \includegraphics[scale=0.1]{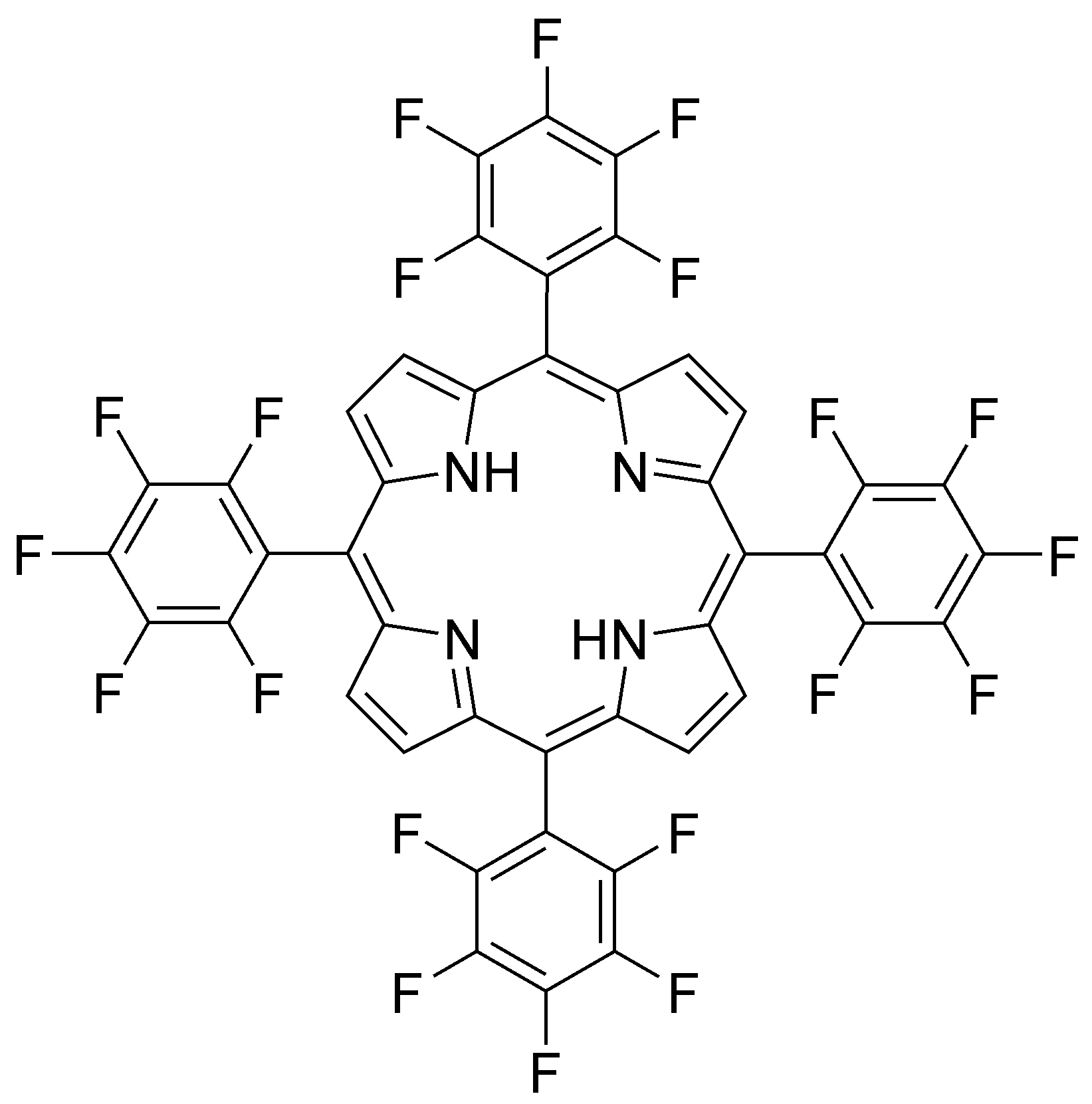}\\ C6F5-H2P &    0.32 &         0.29 &              2.1 &              8.18 &                    \citenum{ref645} \\
    \includegraphics[scale=0.1]{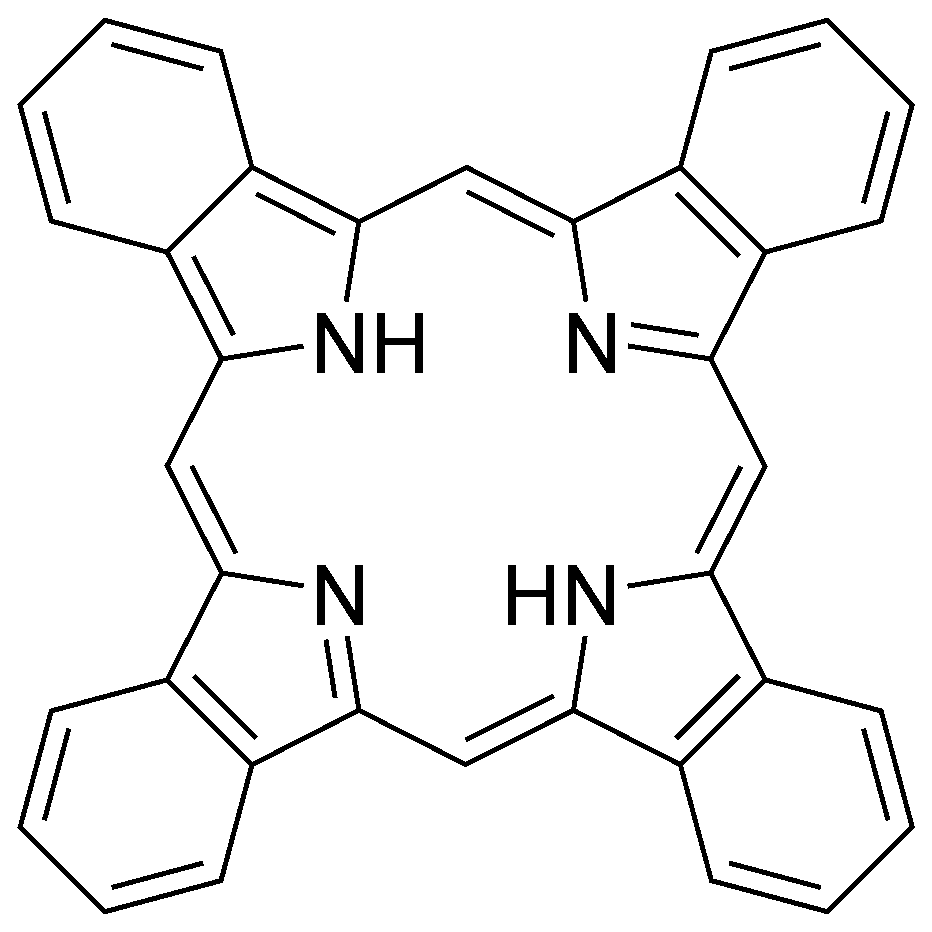}\\ tetrabenzoporphine &     0.17 &         0.09 &             2.2 &            256 &  \citenum{ref611}, \citenum{ref658} \\
    \includegraphics[scale=0.1]{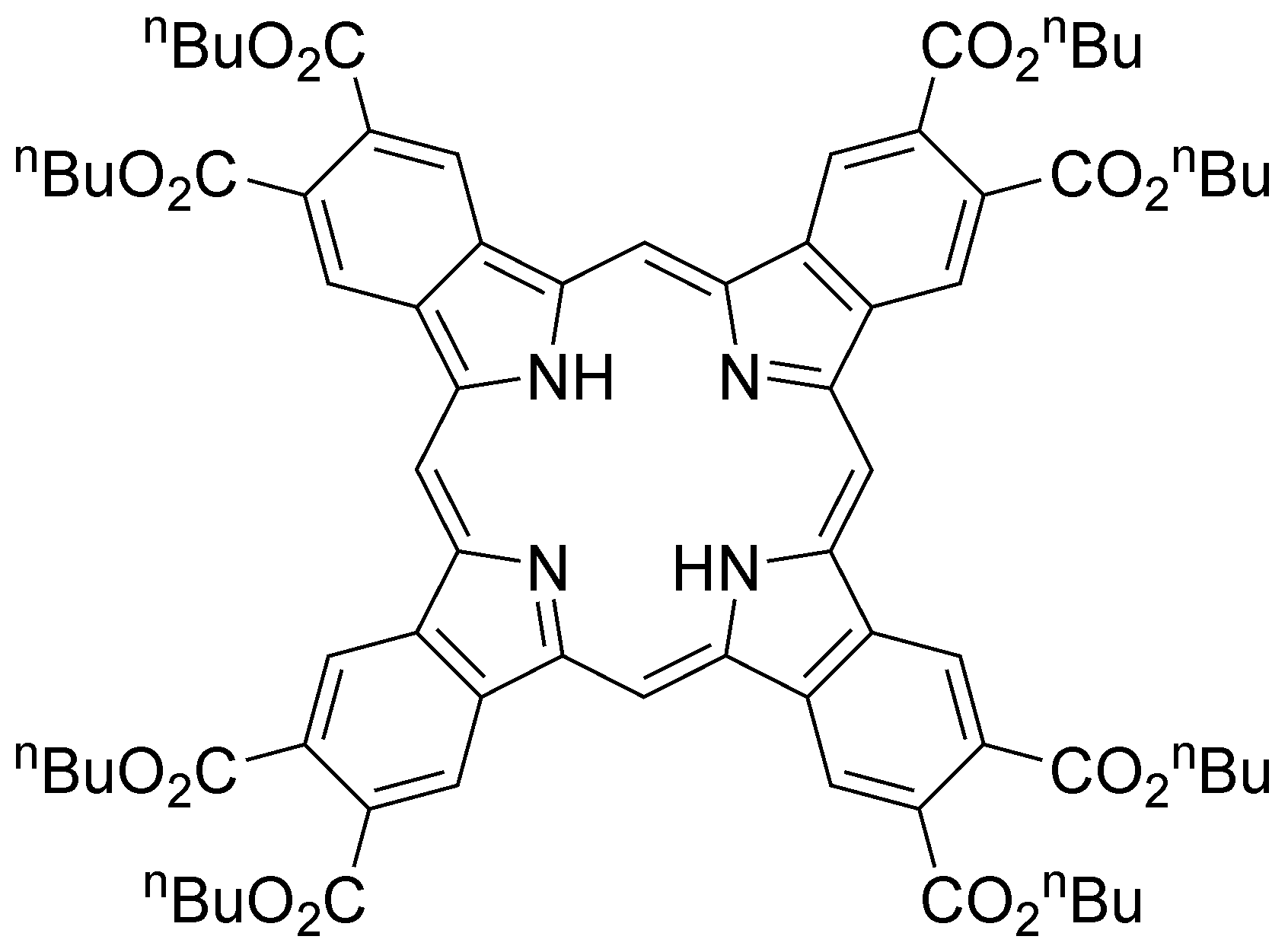}\\ H2TBP(CO2Bu) &     0.14 &         0.11 &              2.1 &            296 &                    \citenum{ref661} \\
    \includegraphics[scale=0.1]{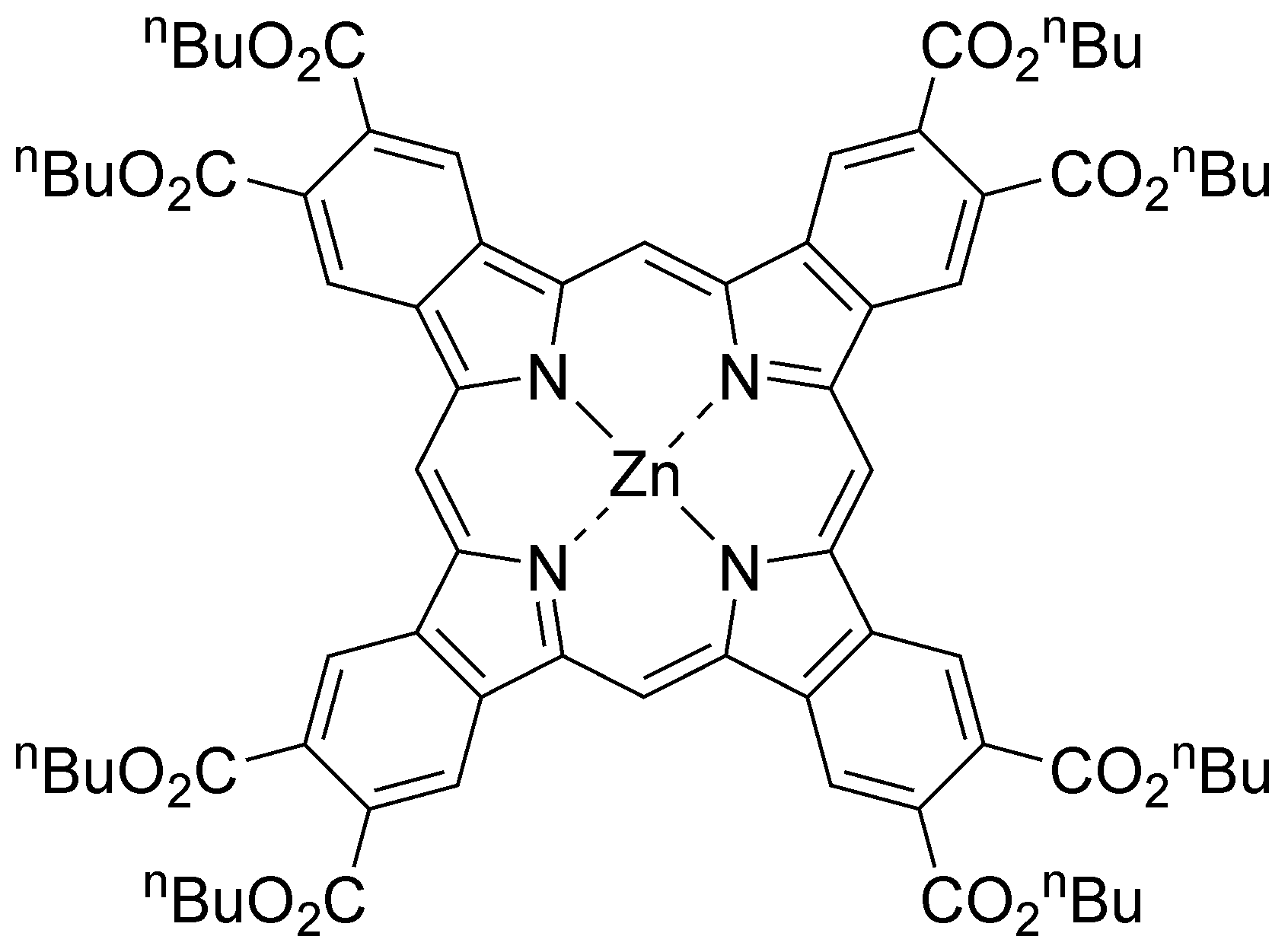}\\ ZnTBP(CO2Bu) &   0.16 &         0.04 &              2.3 &           1760 &                    \citenum{ref661} \\
    \includegraphics[scale=0.1]{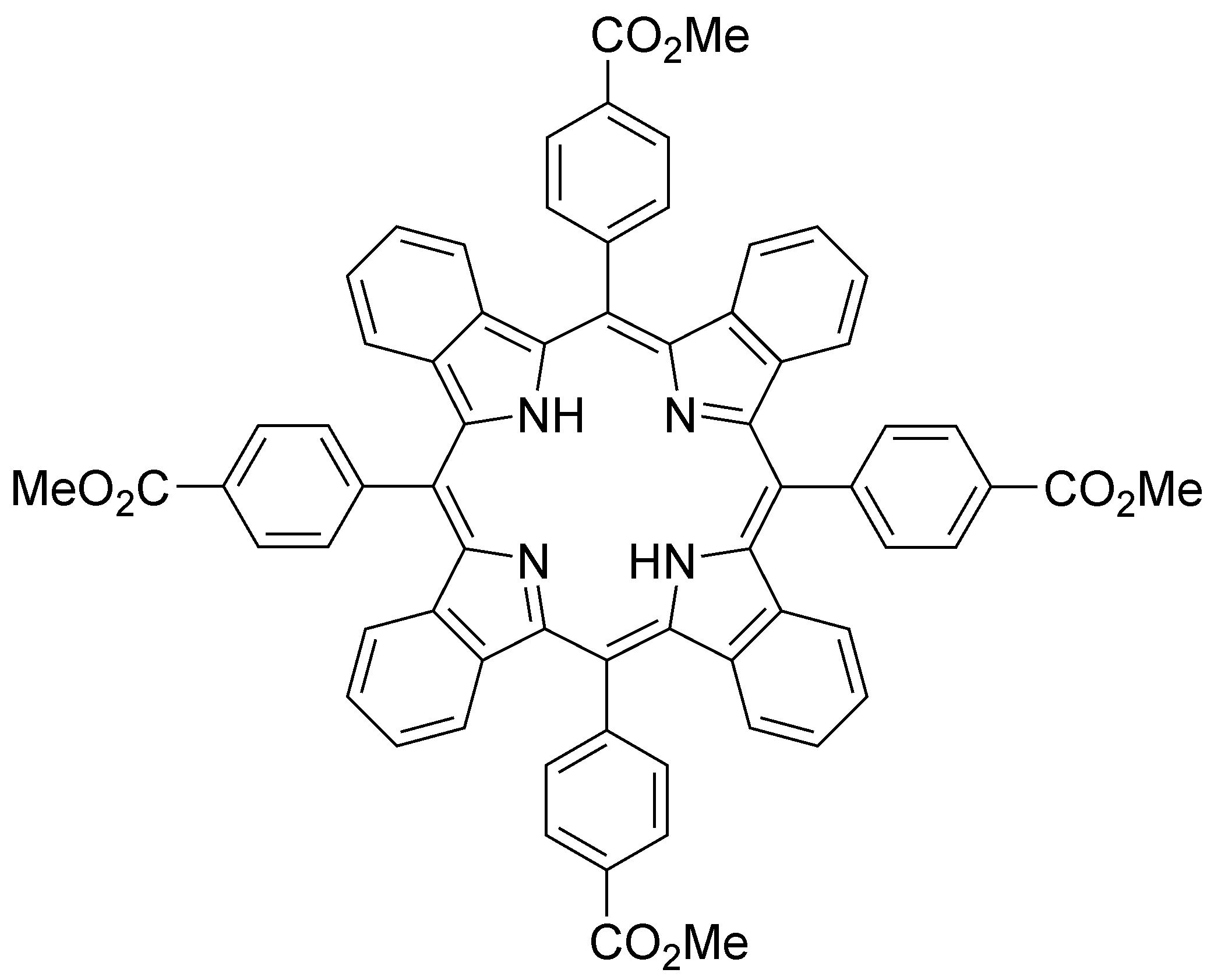} \\ H2TBP(CO2Me)Ph &     0.16 &         0.16 &              2.1 &            103 &                    \citenum{ref662} \\
    \includegraphics[scale=0.1]{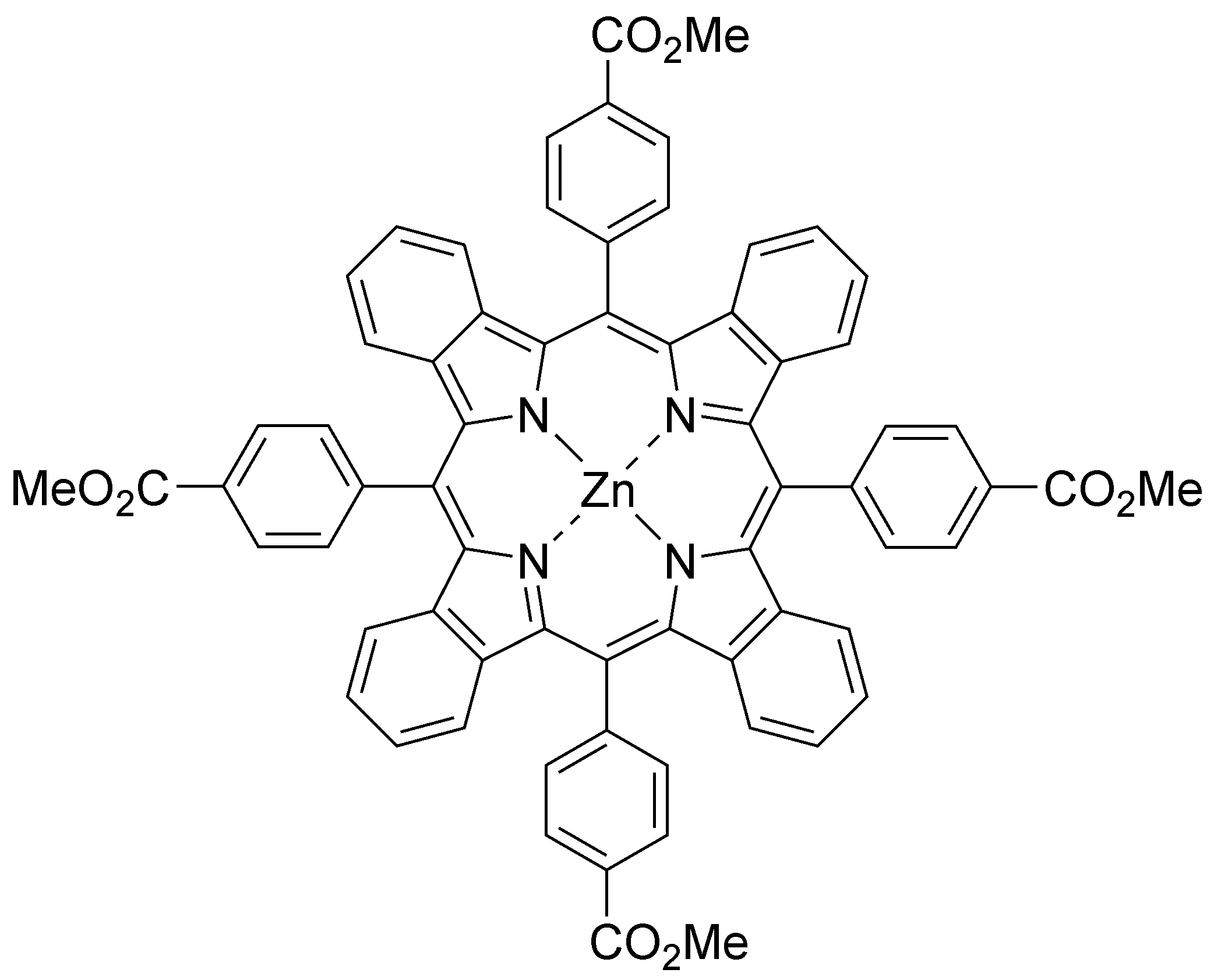}\\  ZnTBP(CO2Me)Ph &    0.13 &         0.07 &              2.2 &            738 &                     \citenum{ref30} \\
    \includegraphics[scale=0.1]{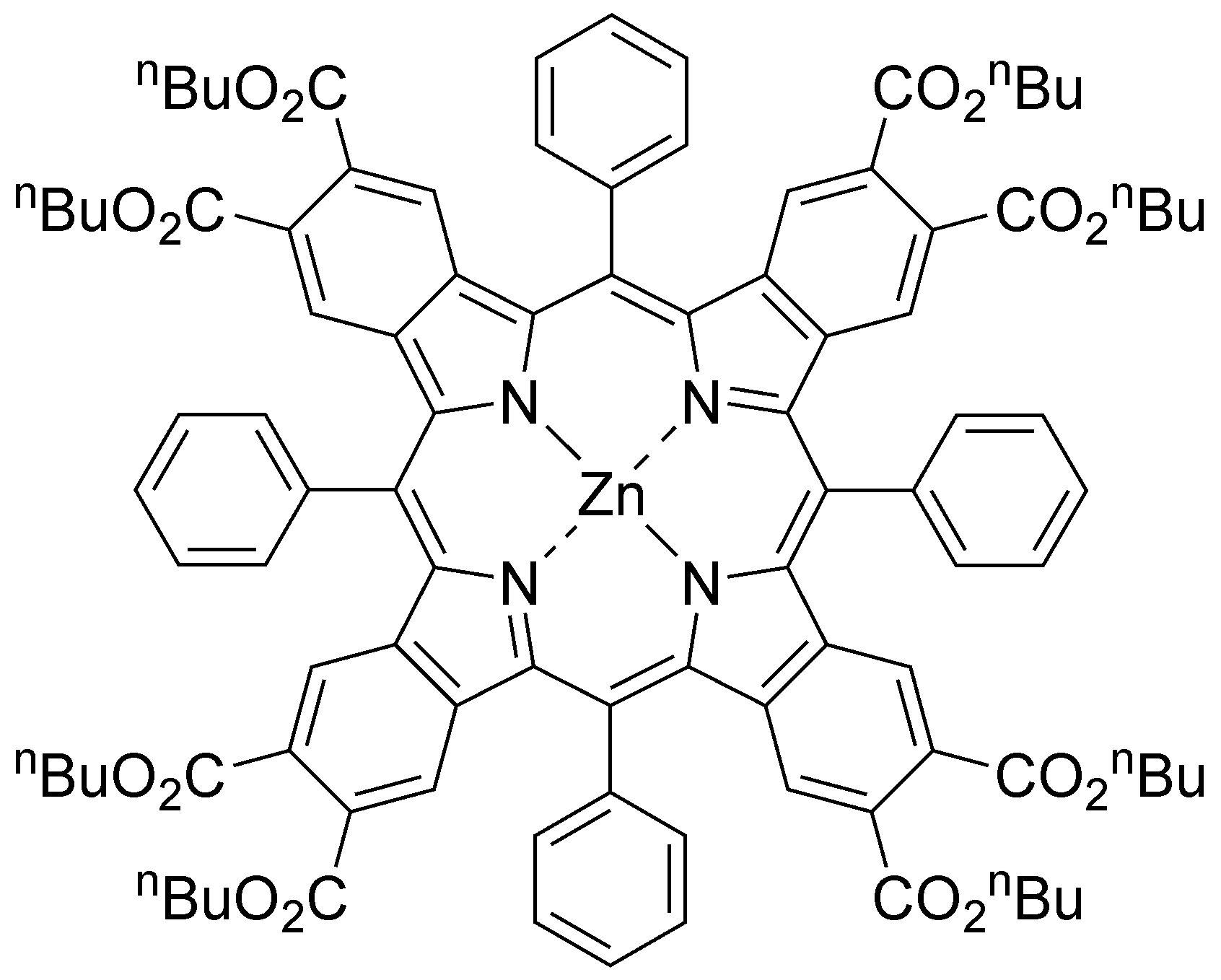} \\  ZnTBP(CO2Bu)Ph &      0.13 &         0.07 &              2.2 &            709 &                    \citenum{ref661} \\
    \includegraphics[scale=0.1]{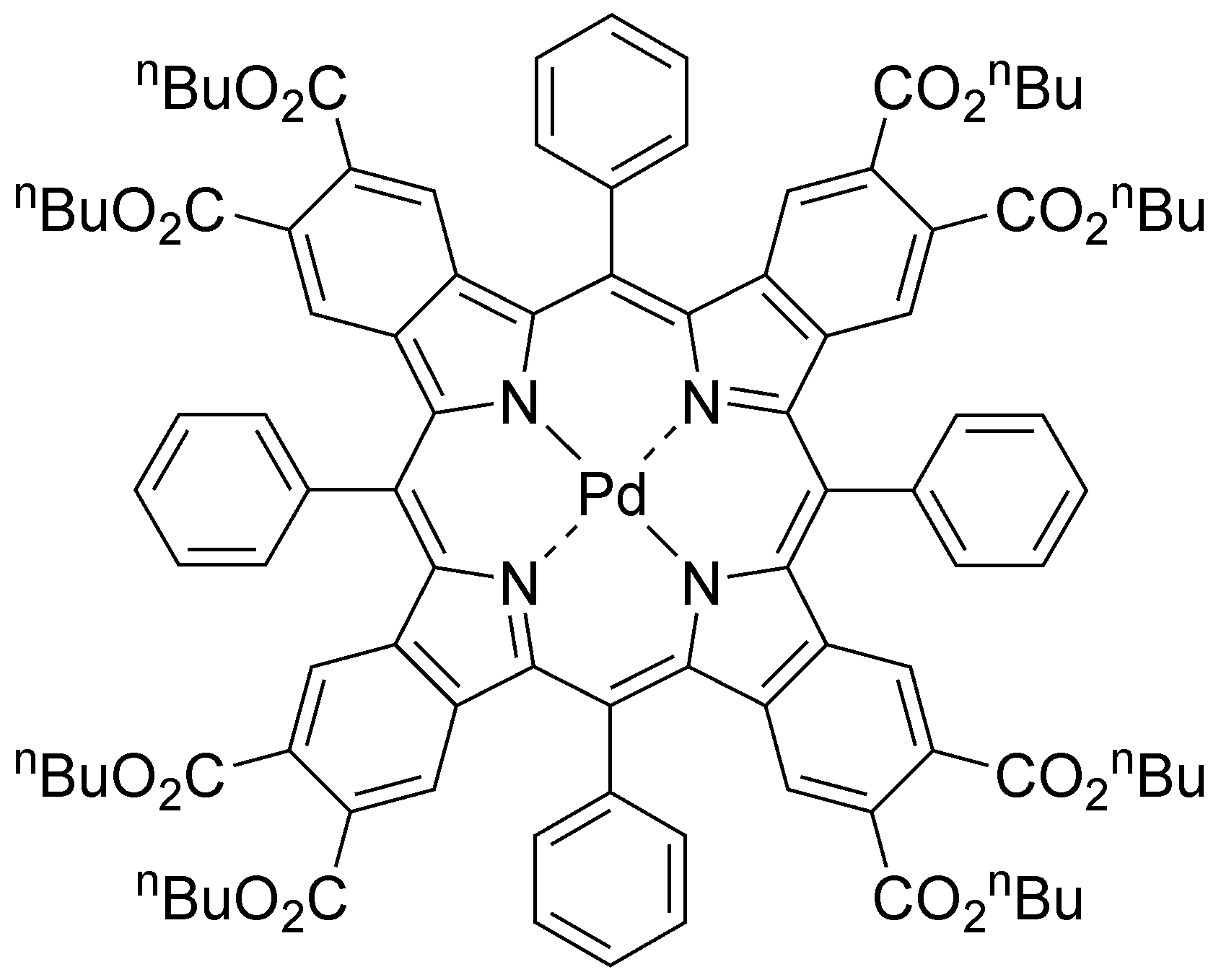}\\ PdTBP(CO2Bu)Ph &     0.16 &         0.21 &              2.2 &             55.7 &                    \citenum{ref661} \\
    \includegraphics[scale=0.1]{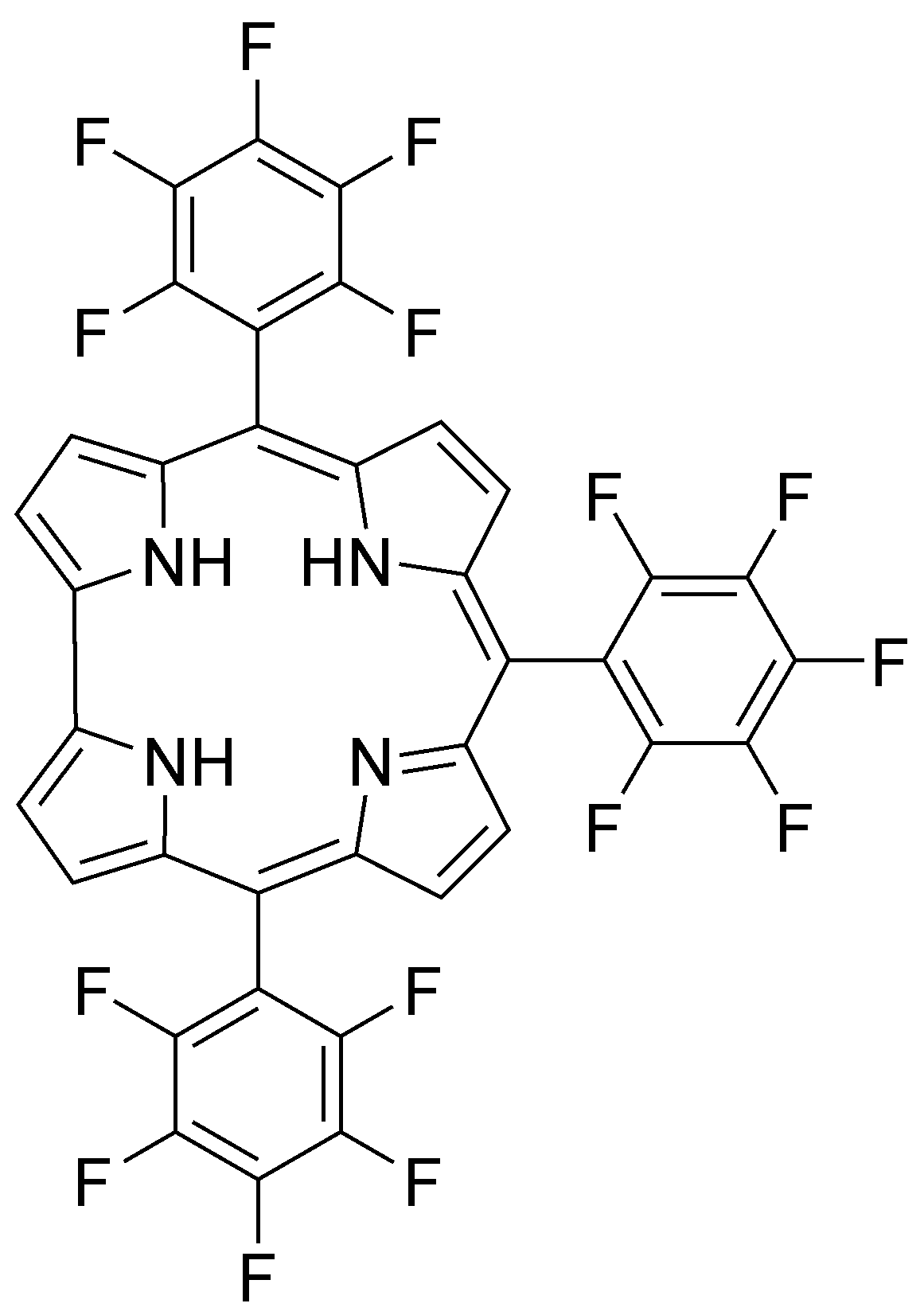}\\ C6F5-Corrole &    0.16 &         0.16 &              2.2 &             98.1 &   \citenum{ref30}, \citenum{ref665} \\
    \includegraphics[scale=0.1]{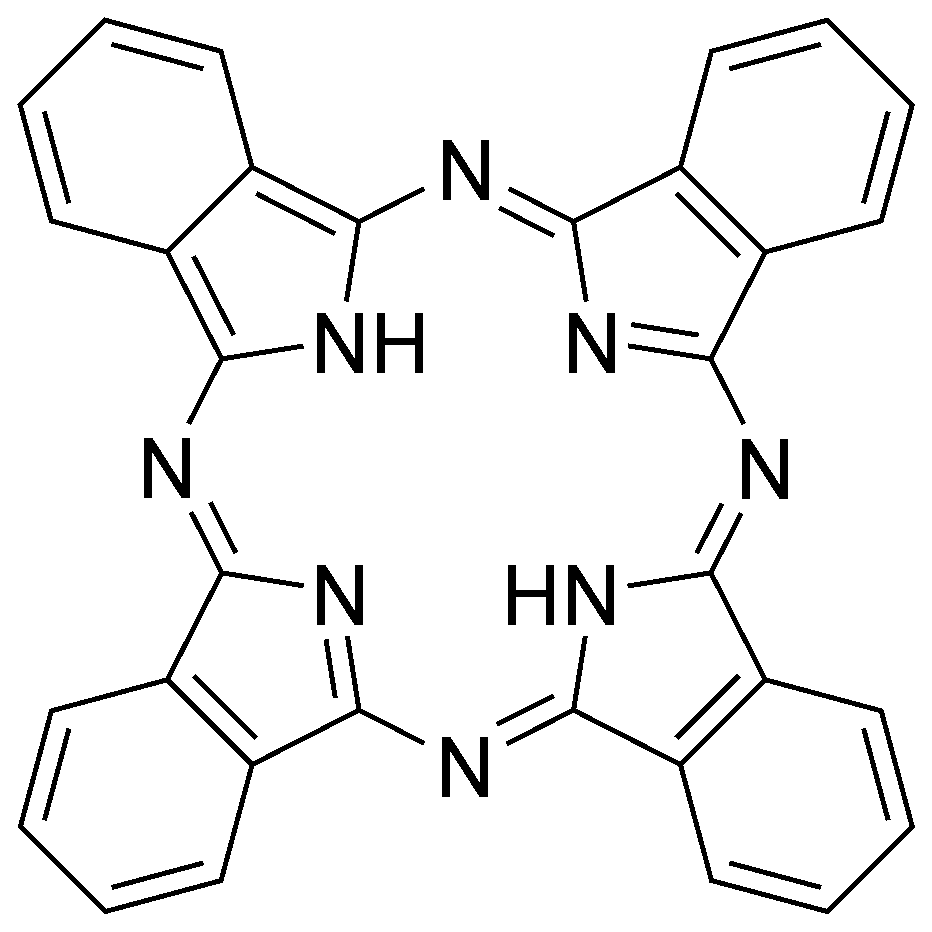}\\ H2Pc &   0.09 &         0.06 &              2.3 &           1797 &  \citenum{ref625}, \citenum{ref676} \\
    \includegraphics[scale=0.1]{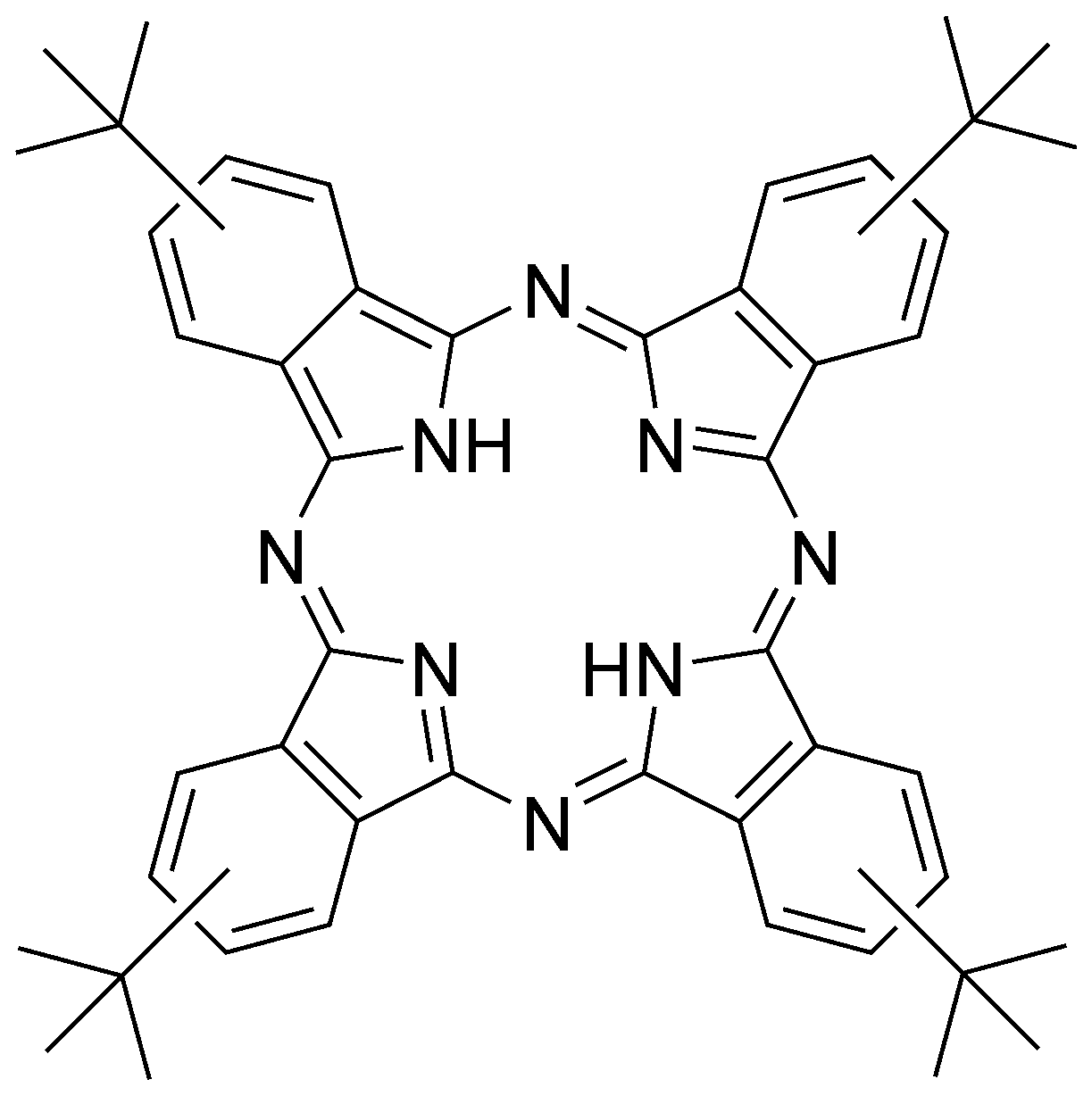} \\ H2Pc(tBu) &     0.10 &         0.06 &              2.3 &           1565 &                    \citenum{ref680} \\
    \includegraphics[scale=0.1]{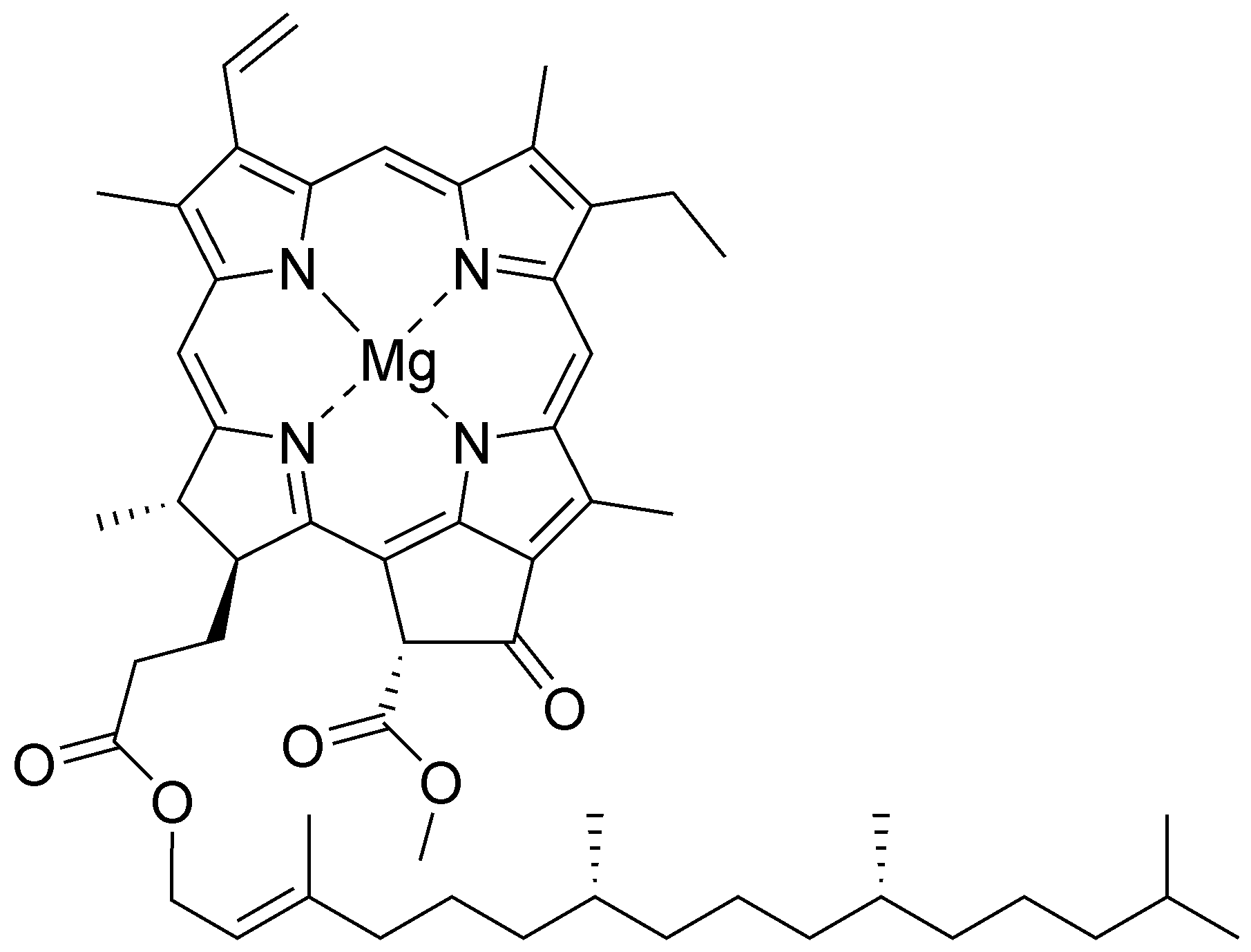}\\ chlorophyll a &     0.14 &         0.06 &              2.2 &            900 &  \citenum{ref687}, \citenum{ref688} \\
    \includegraphics[scale=0.1]{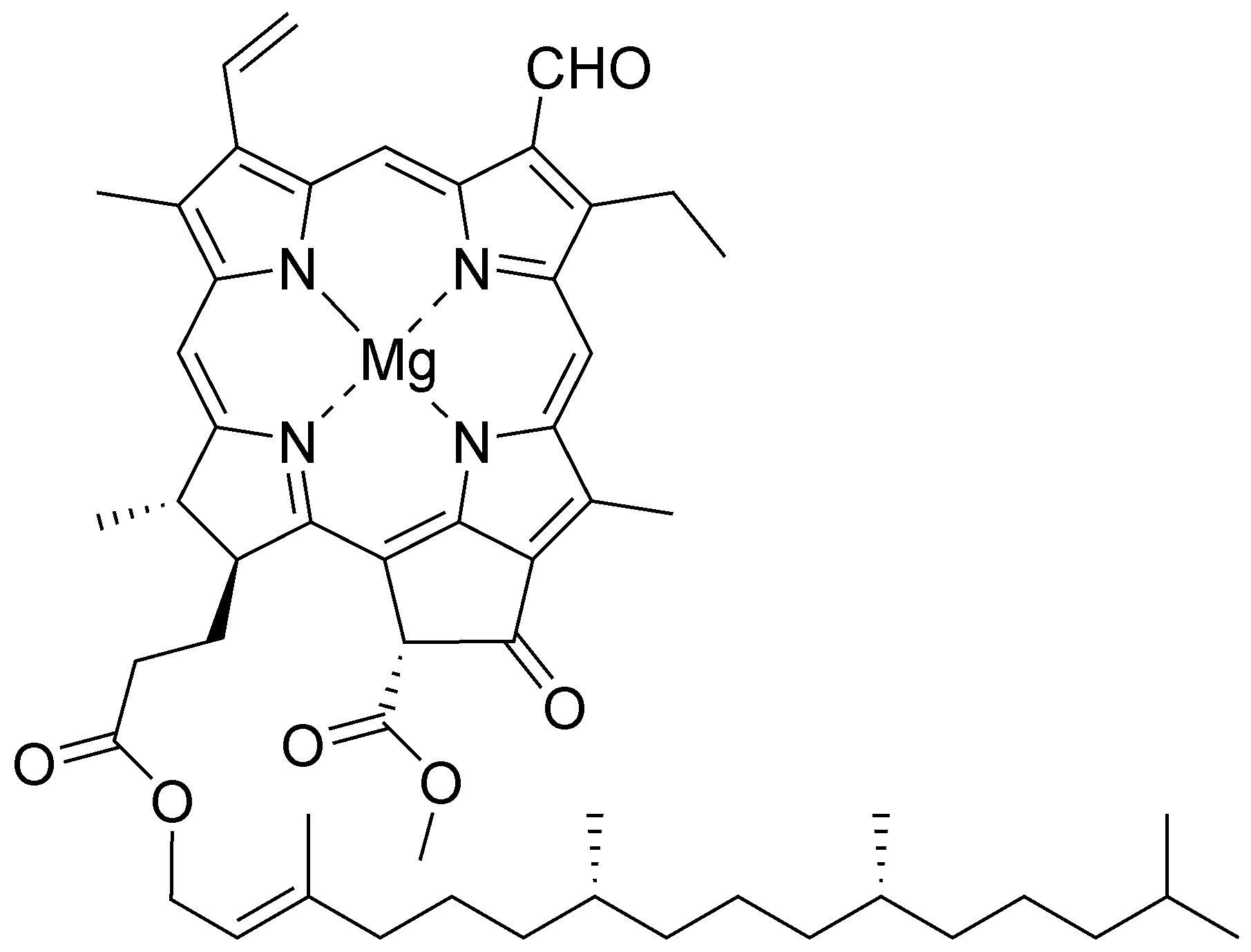} \\ chlorophyll b &    0.16 &         0.03 &              2.3 &           2081 &  \citenum{ref688}, \citenum{ref691} \\
    \includegraphics[scale=0.1]{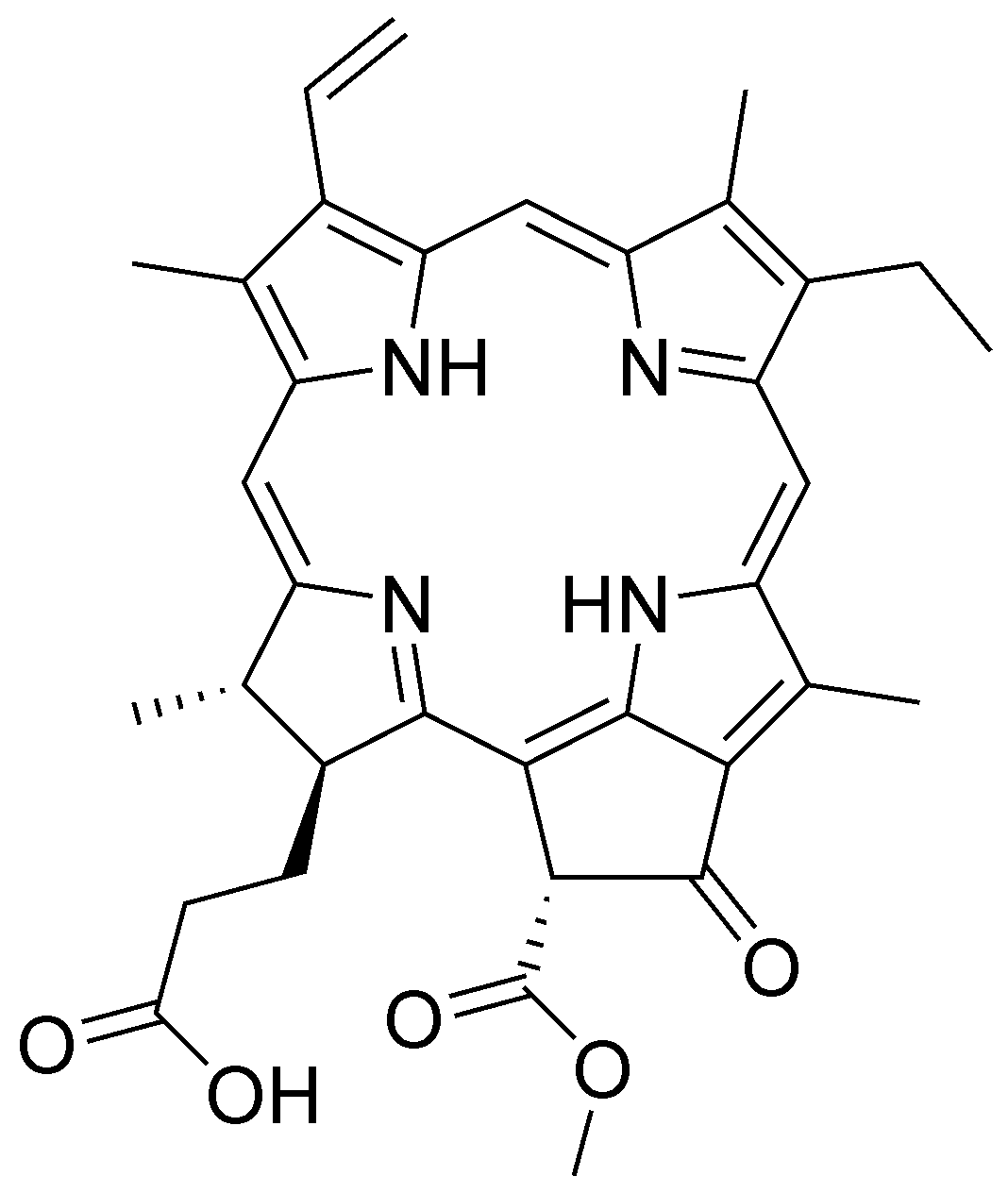}\\ pheophorbide a &     0.17 &         0.04 &              2.3 &           1438 &                    \citenum{ref693} \\
    \includegraphics[scale=0.1]{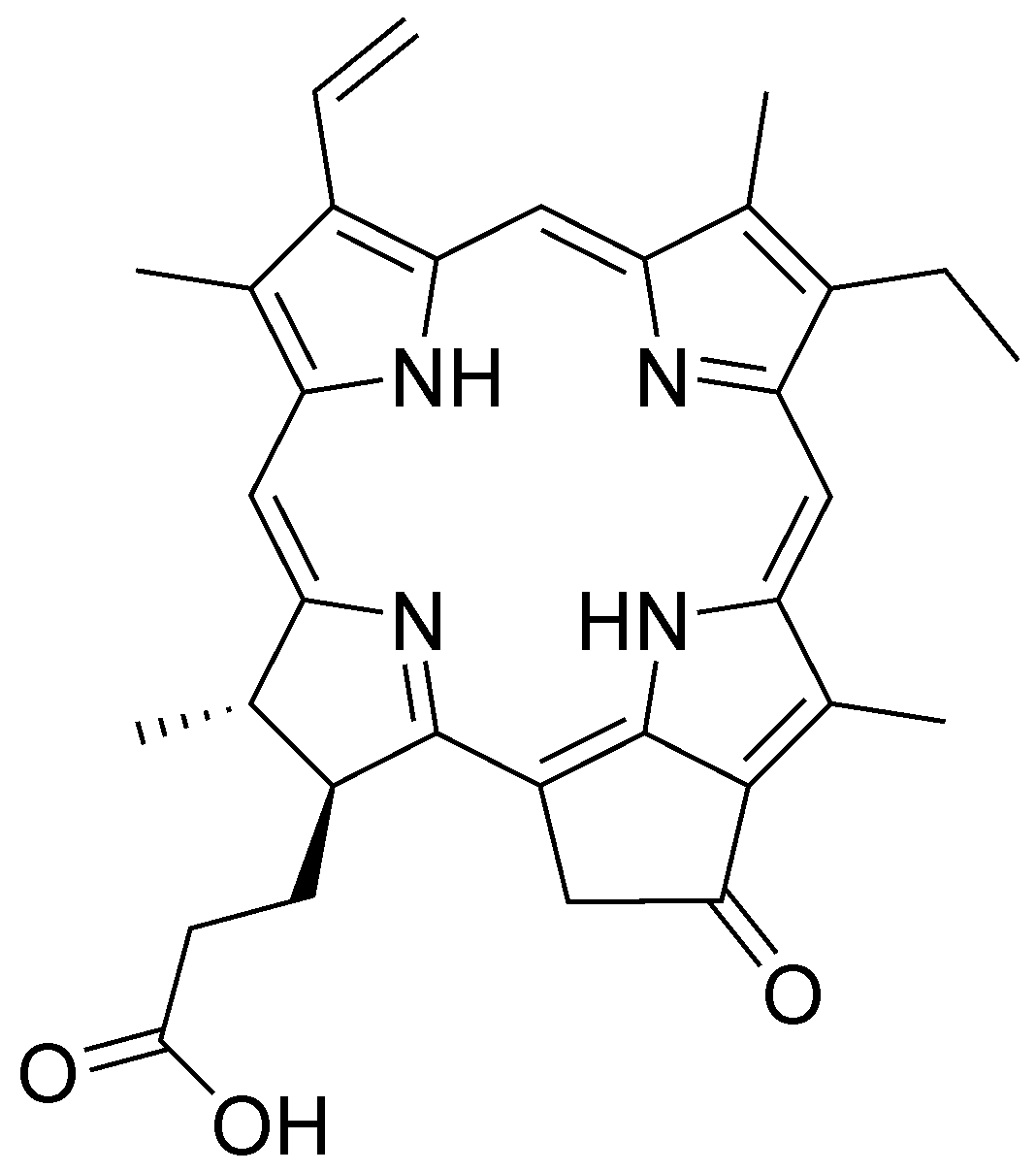}\\ pyropheophorbide a &        0.17 &         0.04 &              2.3 &           1669 &  \citenum{ref696}, \citenum{ref697} \\
    \includegraphics[scale=0.1]{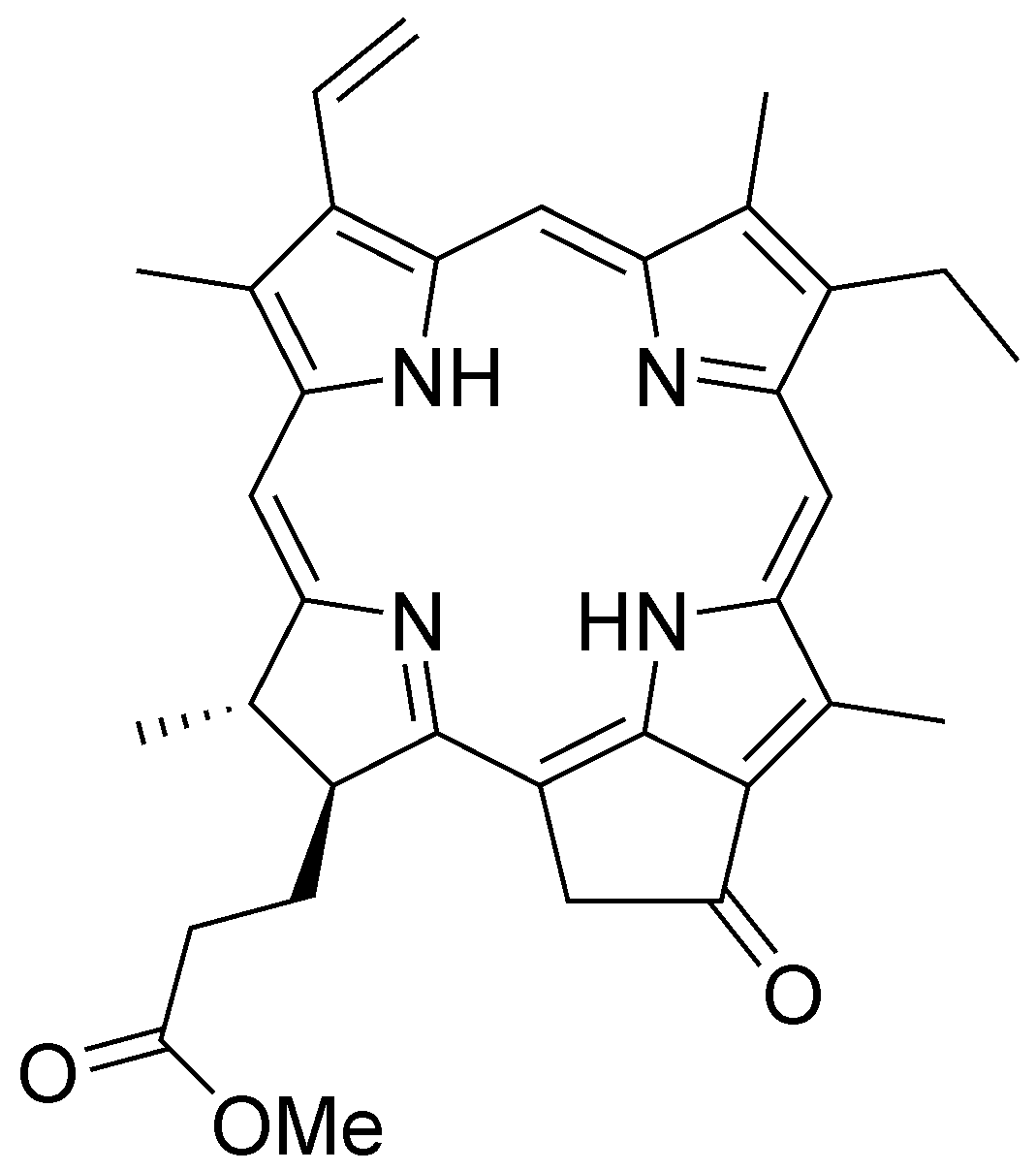}\\  pyropheophorbide a methyl ester &     0.18 &         0.04 &              2.3 &           1,292 &  \citenum{ref698}, \citenum{ref699} \\
    \includegraphics[scale=0.1]{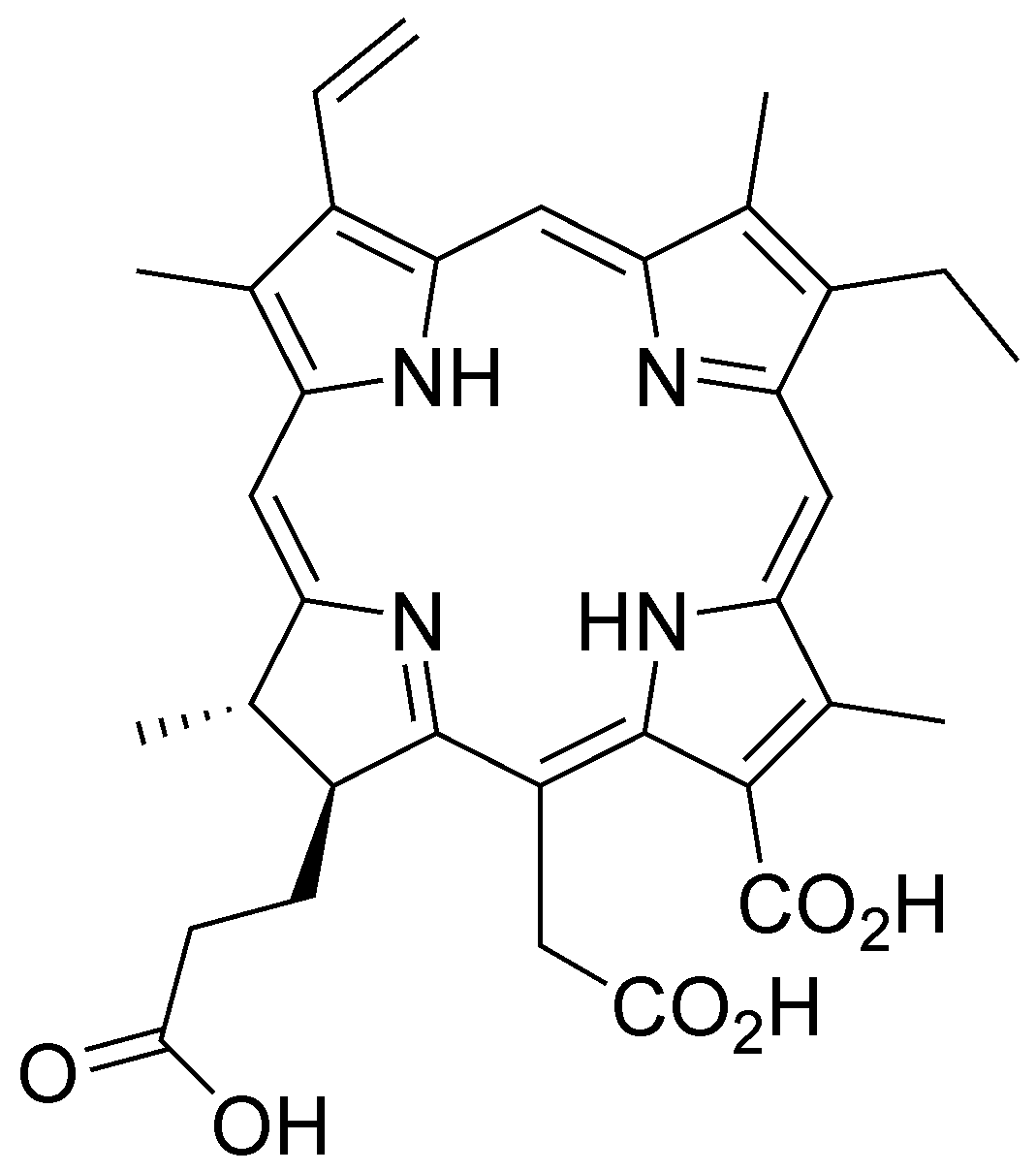}\\ chlorin e6 &    0.17 &         0.03 &              2.4 &           3017934 &  \citenum{ref704}, \citenum{ref705} \\ 
    \includegraphics[scale=0.1]{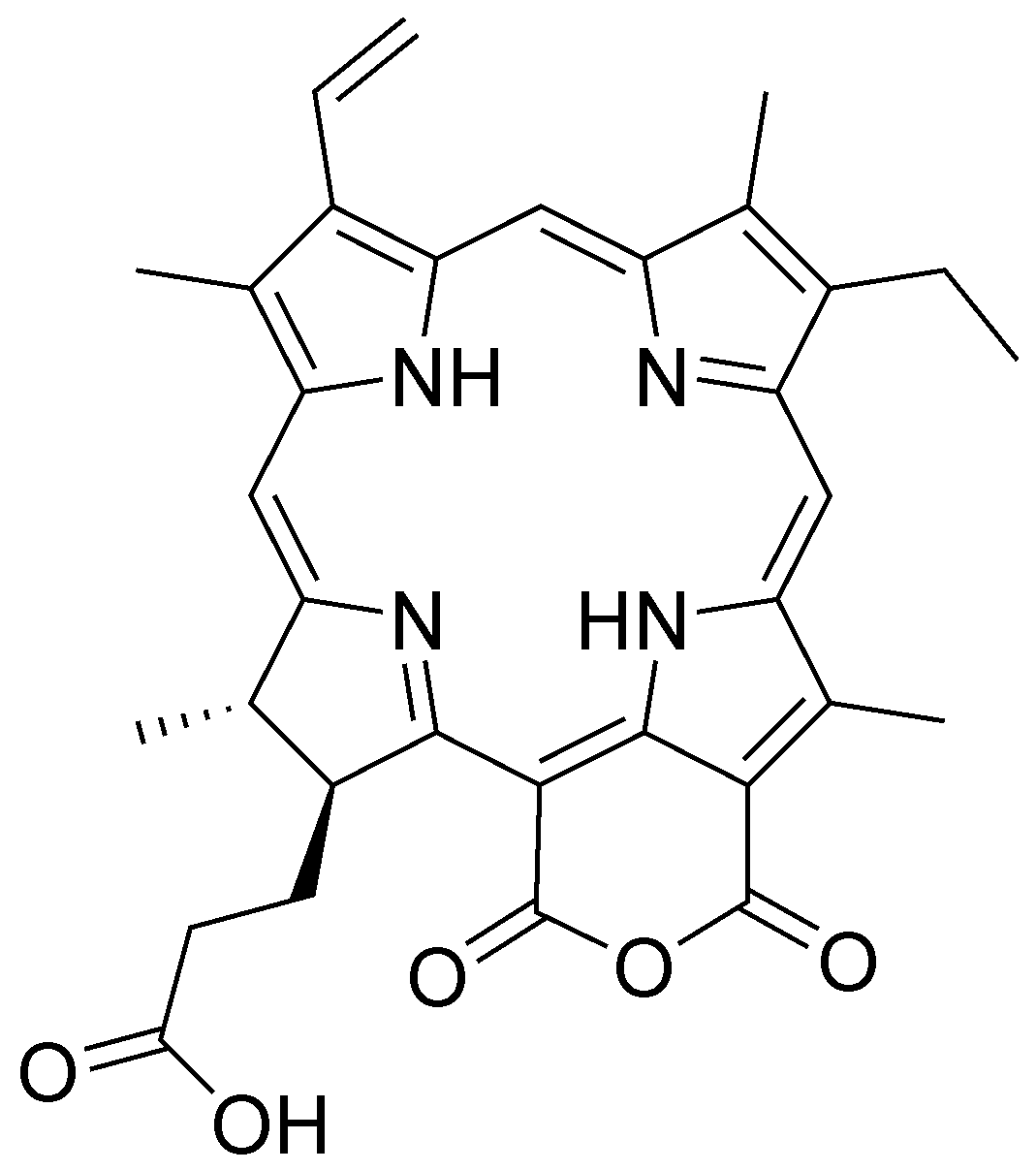}\\ purpurin 18 &        0.16 &         0.03 &              2.4 &           2,588 &  \citenum{ref706}, \citenum{ref707} \\ 
    \includegraphics[scale=0.1]{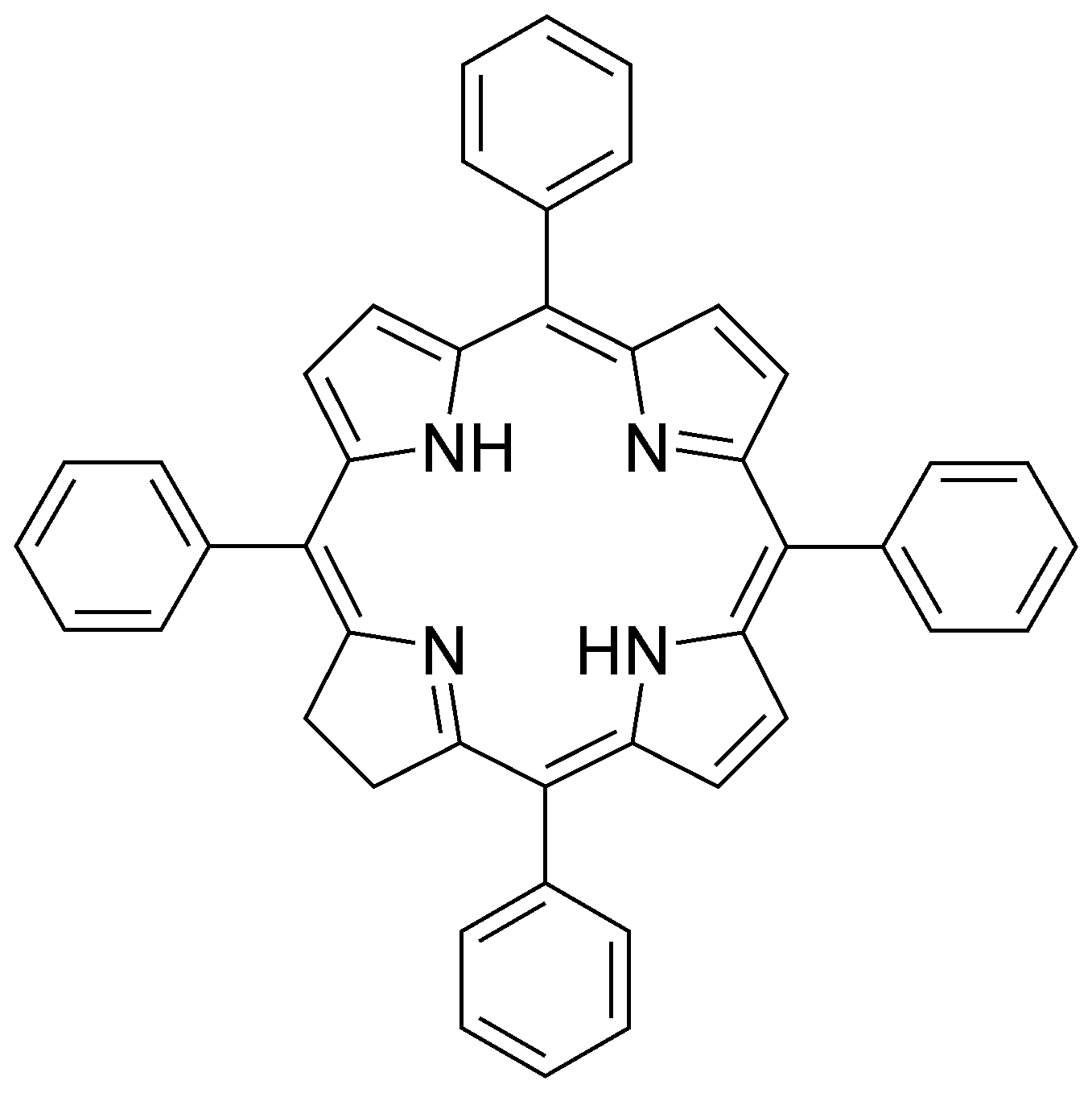} \\ H2TPC &    0.17 &         0.03 &              2.4 &           2,556 &  \citenum{ref709}, \citenum{ref710} \\ \pagebreak
    \includegraphics[scale=0.1]{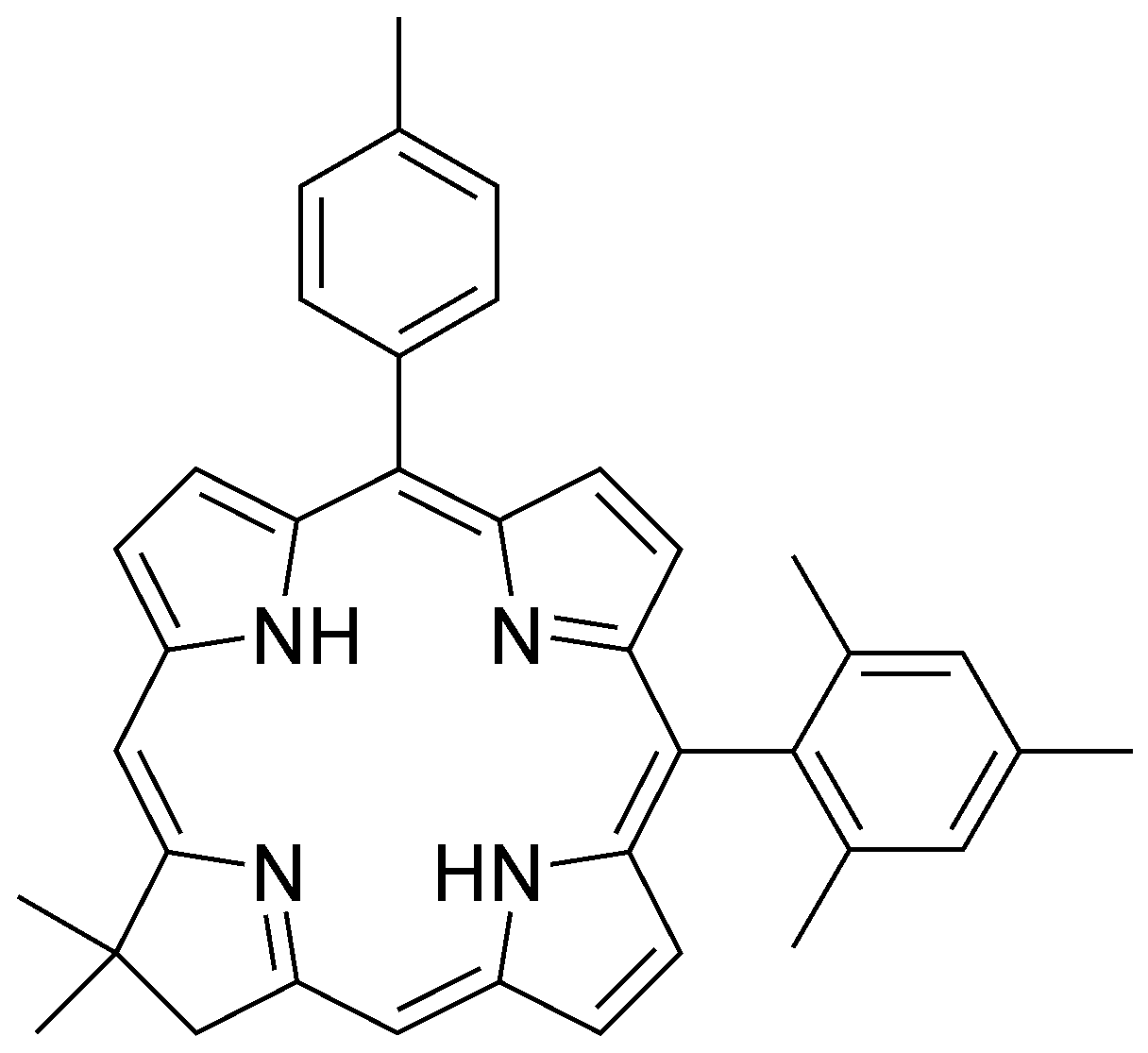}\\ H2C-1 &    0.17 &         0.03 &              2.3 &           2,241 &  \citenum{ref716}, \citenum{ref717} \\
    \includegraphics[scale=0.1]{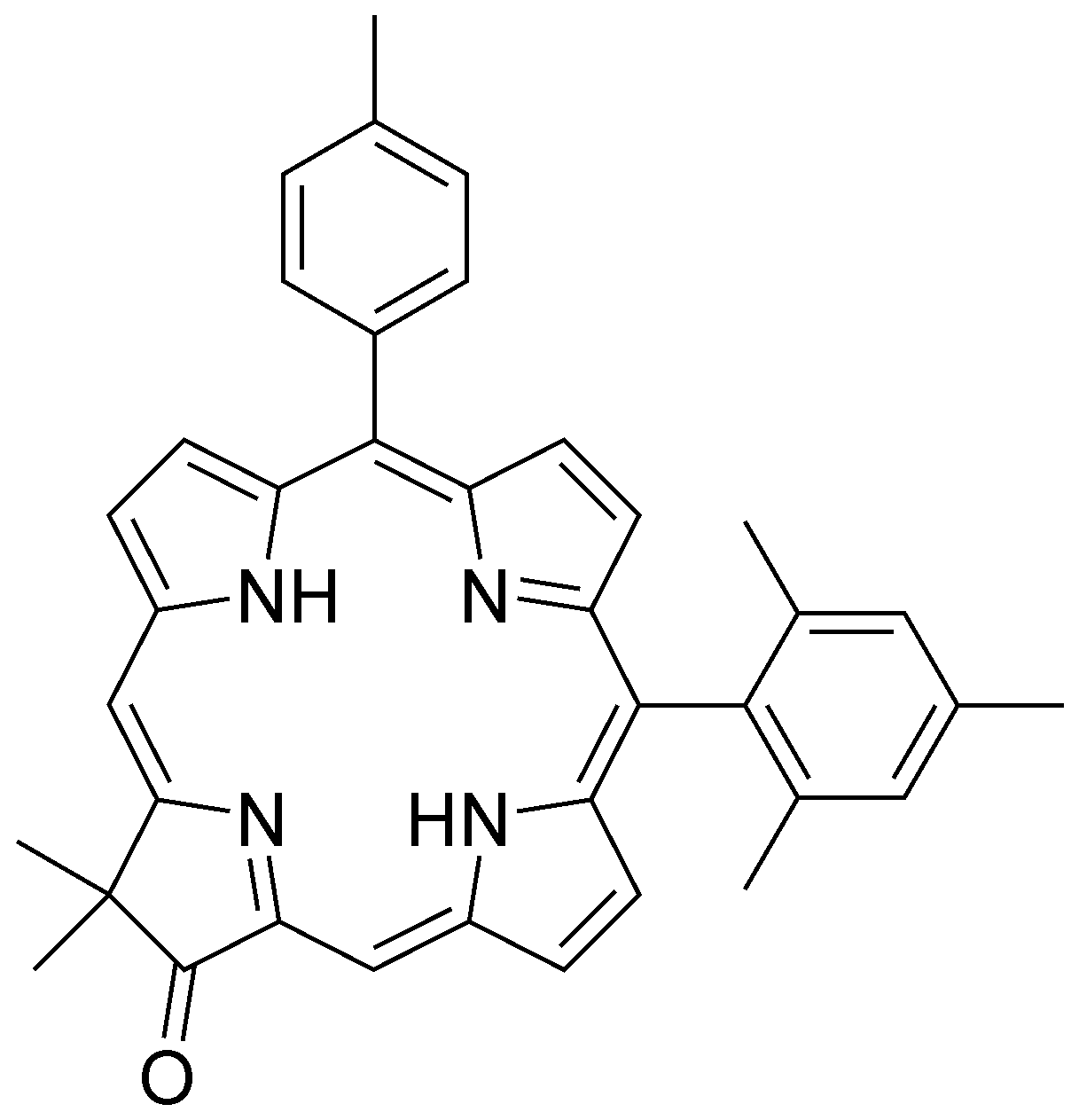}\\ H2COxo-1 &     0.19 &         0.05 &              2.2 &            860 &                    \citenum{ref717} \\
    \includegraphics[scale=0.1]{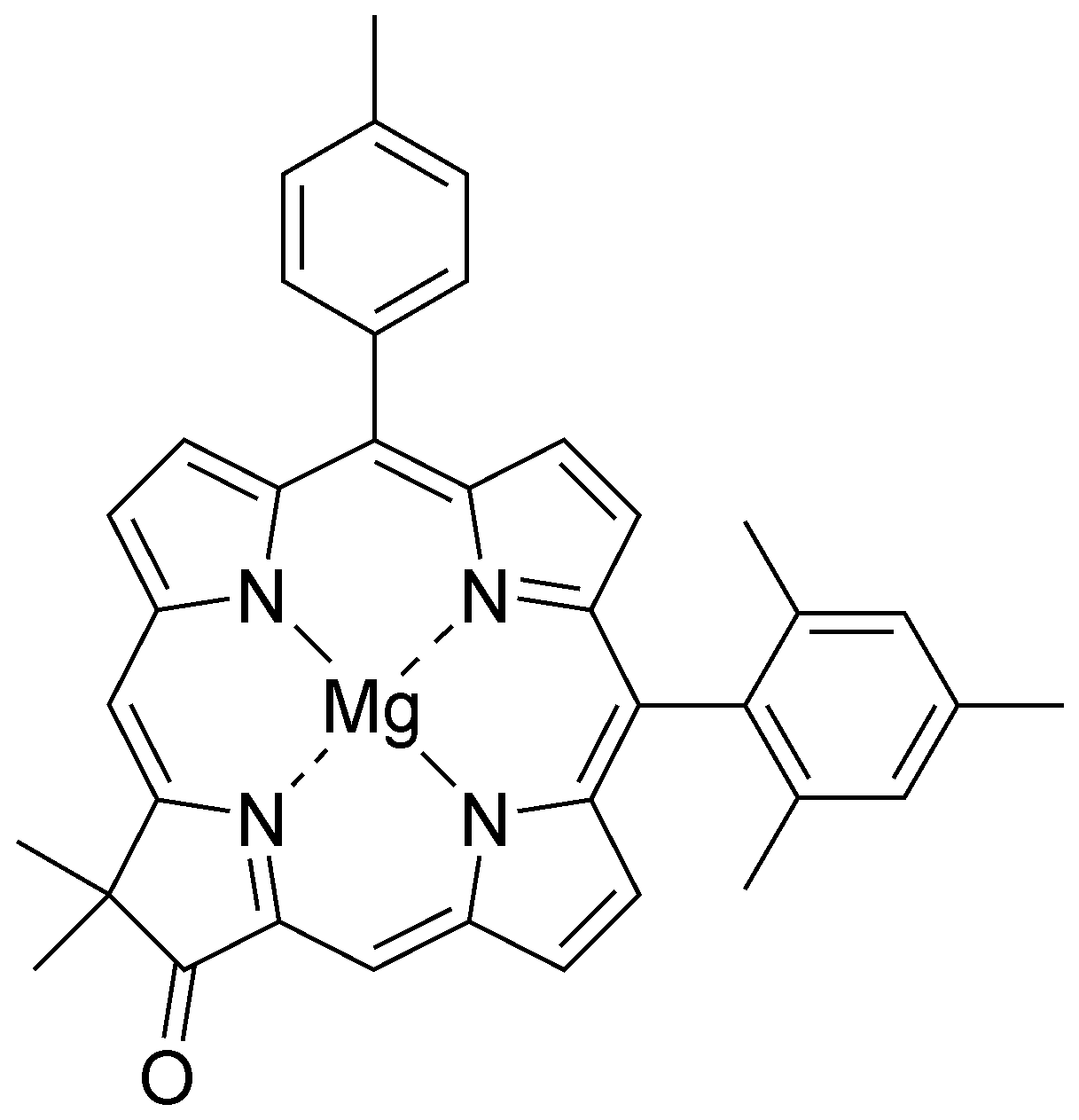} \\ MgCOxo-1 &    0.18 &         0.09 &              2.2 &            279 &                    \citenum{ref717} \\
    \includegraphics[scale=0.1]{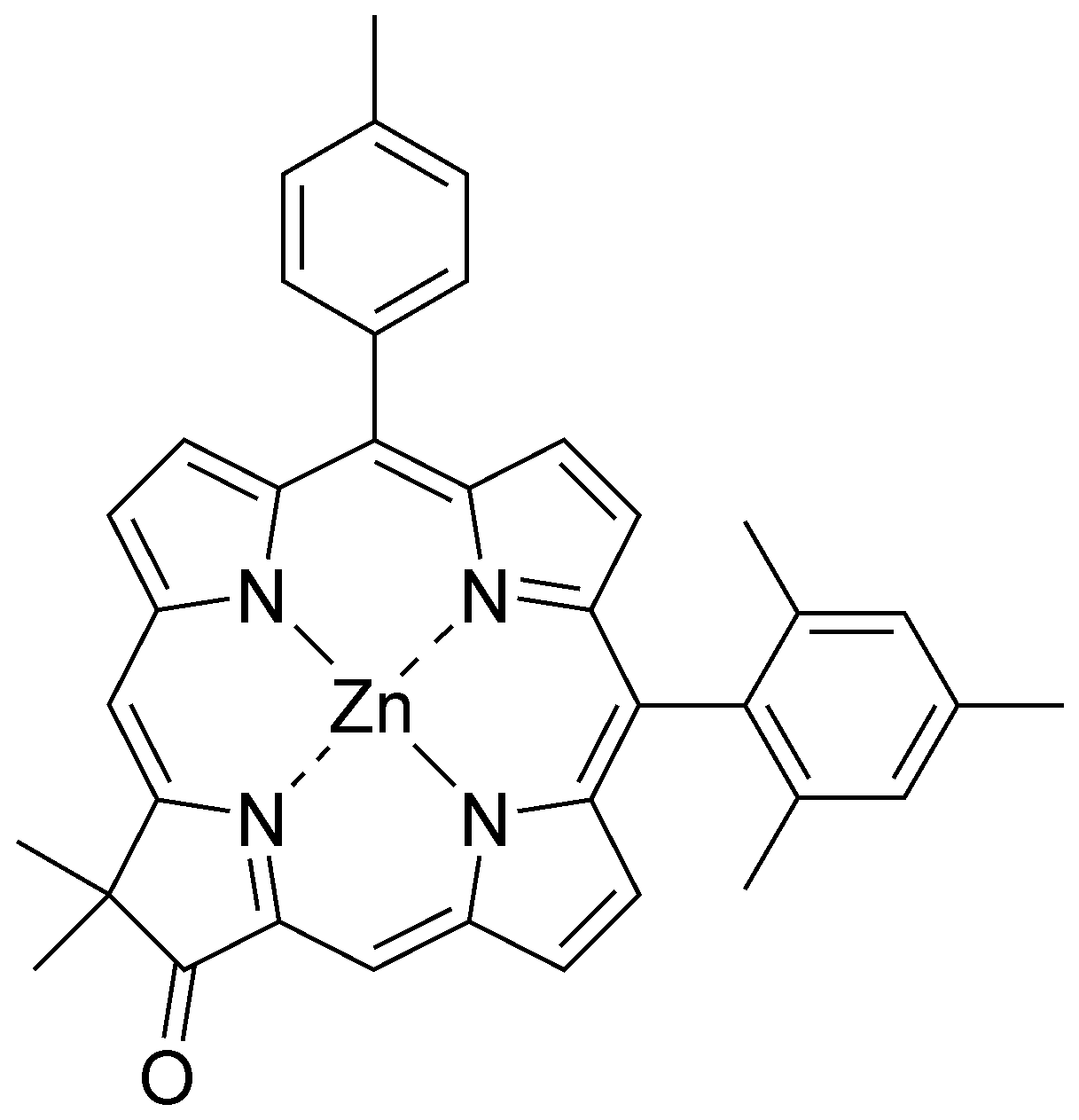}\\ ZnCOxo-1 &          0.17 &         0.06 &              2.2 &            676 &                    \citenum{ref717} \\

    \end{longtable}
    
    \end{widetext}

\end{document}